\documentclass[%
 reprint,
superscriptaddress,
 amsmath,amssymb,
 aps,
pra,
floatfix,
]{revtex4-1}

\usepackage{graphicx}
\usepackage{dcolumn}
\usepackage{bm}
\usepackage{siunitx}
    \DeclareSIUnit \counts{Cts}
    \DeclareSIUnit \electron{e}
    \DeclareSIUnit \eV{eV}
    \DeclareSIUnit \molar{M}
    \DeclareSIUnit \Gauss{G}
\usepackage{braket}
\newcommand{\ra}[1]{\renewcommand{\arraystretch}{#1}}

\usepackage[colorlinks=true, allcolors=blue]{hyperref}

\begin{document}

\preprint{}

\title{Spin Readout Techniques of the Nitrogen-Vacancy Center in Diamond}
\author{David A. Hopper}
\affiliation{Quantum Engineering Laboratory, Department of Electrical and Systems Engineering, University of Pennsylvania, Philadelphia Pennsylvania 19104, USA}
\affiliation{Department of Physics and Astronomy, University of Pennsylvania, Philadelphia Pennsylvania 19104, USA}

\author{Henry J. Shulevitz}
\affiliation{Quantum Engineering Laboratory, Department of Electrical and Systems Engineering, University of Pennsylvania, Philadelphia Pennsylvania 19104, USA}

\author{Lee C. Bassett}
\email{lbassett@seas.upenn.edu}
\affiliation{Quantum Engineering Laboratory, Department of Electrical and Systems Engineering, University of Pennsylvania, Philadelphia Pennsylvania 19104, USA}

\date{\today}

\begin{abstract}
The diamond nitrogen-vacancy (NV) center is a leading platform for quantum information science due to its optical addressability and room-temperature spin coherence. However, measurements of the NV center's spin state typically require averaging over many cycles to overcome noise. Here, we review several approaches to improve the readout performance and highlight future avenues of research that could enable single-shot electron-spin readout at room temperature.
\end{abstract}

\maketitle

\begin{quote}
\centering
``The only thing you can do easily is be wrong, and that's hardly worth the effort.''
\end{quote}
\begin{flushright}
 ---Norton Juster, \emph{The Phantom Tollbooth}
\end{flushright}
\section{Introduction}

 Recent, rapid advances in creating, detecting, and controlling quantum-mechanical states in engineered systems heralds the beginning of the quantum-information era.
 A diverse set of physical platforms, including superconducting circuits \cite{Gambetta2017}, cold ions \cite{Brown2016}, integrated photonics \cite{Silverstone2016}, and spins in semiconductors \cite{Awschalom2013}, have enabled progress toward fault-tolerant quantum computation, quantum-secure communication systems, and unparalleled sensing technologies.
Nevertheless, most platforms remain in the early engineering stages and face substantial technical challenges.
 A common challenge, and critical criterion for scalable quantum information processing \cite{DiVincenzo2000}, is reliably measuring the quantum state.
 The issue of precision measurement is one of the oldest and most subtle aspects of quantum theory -- and arguably the most essential for many practical applications.
 Several authors have reviewed general considerations for quantum measurements \cite{nielsen2000, Clerk2010}.
 Here, we focus on the problem as applied to the nitrogen-vacancy (NV) center in diamond, which has emerged as a compelling solid-state qubit for a wide range of quantum technologies.

 Point defects in wide-bandgap semiconductors are analogous to molecules trapped within a crystalline host.
 A small subset of these point defects functions as qubits with optical addressability, exceptional spin coherence properties, and room-temperature operation \cite{Heremans2016}.
 The diamond nitrogen-vacancy (NV) center is the prototypical defect spin qubit, and the most intensely studied \cite{Doherty2013}.
 A truly versatile platform, \mbox{the NV} center has been utilized for designing quantum memories \cite{Dutt2007,Maurer2012,Pfender2017a}; addressing individual nuclear spins \cite{Childress2006,Neumann2010,Liu2017}; engineering nanoscale sensors of magnetism \cite{Casola2018}, proteins \cite{Lovchinsky2016}, and \mbox{chemicals \cite{Aslam2017}}; exploring hybrid quantum mechanical systems \cite{Arcizet2011}; and testing the fundamental principles of quantum mechanics through loophole free violations of Bell's inequality \cite{Hensen2015}.
 In the course of these investigations, several techniques have been developed to measure the NV center's spin state that offer certain advantages for specific circumstances.
 Here, we review the leading techniques for NV spin readout by presenting the physical mechanisms, discussing the state-of-the-art and considering the potential for further improvement.
 Due to the breadth of NV research, we direct readers to detailed reviews on quantum sensing \cite{Degen2017}, NV magnetometry \cite{Rondin2014}, nanodiamond sensing \cite{Schirhagl2014}, and nanophotonics in diamond \cite{Schroder:16} for an overview of these application areas.

 The review is organized as follows:
 Section~\ref{sec:readout_performance} overviews several spin-readout performance metrics commonly used in the community; Section~\ref{sec:traditional_readout} introduces the traditional approach to spin readout using photoluminescence (PL);
 Section~\ref{sec:collection_efficiency} discusses recent efforts to improve photon collection efficiency;
 Section~\ref{sec:radiative_lifetime} considers how altering the excited state lifetime affects spin readout;
 Section~\ref{sec:low_temp_readout} introduces the resonance-fluorescence technique for single-shot spin readout at low temperature;
 Section~\ref{sec:nuclear_assisted} describes how coupled nuclear spins can improve the electron-spin readout;
 Section~\ref{sec:scc} overviews protocols for spin-to-charge conversion;
 Section~\ref{sec:photocurrent} discusses recent advances in measuring the spin state through photocurrent;
 Section~\ref{sec:measurement_overhead} explains how accounting for measurement overhead can improve the time-averaged signal-to-noise ratio and sensitivity;
 Section~\ref{sec:signal_processing} discusses the use of real-time signal processing;
 Section~\ref{sec:discussion} considers the potential for combining different techniques;
 Section~\ref{sec:conclusion} summarizes the review and provides an outlook on the future of NV applications enabled by maturing readout techniques.

\section{Quantifying Readout Performance \label{sec:readout_performance}}

 Various metrics are used by the NV community to quantify readout performance, each with intuitive advantages for specific applications.
 As we show in this section, the common metrics all relate to the signal-to-noise ratio (SNR) of the measurement, which provides a useful basis to compare different readout techniques.
 We consider projective measurements where the goal is to distinguish between two quantum states, $\ket{0}$ and $\ket{1}$.
 Therefore, we define the SNR for a differential measurement,
\begin{equation}
\textrm{SNR} = \frac{\langle S_0 \rangle - \langle S_1 \rangle}{\sqrt{\sigma_0^2 + \sigma_1^2}},
\label{eqn:diff_snr}
\end{equation}
where $\langle S_i \rangle$ is the mean signal for a single measurement of spin state $\ket{i}$, and $\sigma_i$ is the associated noise.
 Classical signal processing \cite{McDonough1995} and superconducting qubits \cite{Vijay2011} both employ an analogous definition of differential SNR.
 In the following subsections, we discuss common optical-detection signals and their associated SNR, relate the SNR to other spin-readout metrics, and discuss how to include averaging over multiple experimental cycles.

\subsection{Photon Summation}
 In many situations, the signal is simply the number of photons detected in a fixed readout cycle.
 In this case, Equation~(\ref{eqn:diff_snr}) takes the form:
\begin{equation}
\textrm{SNR} = \frac{\alpha_0 - \alpha_1}{\sqrt{\alpha_0 + \alpha_1}},
\label{eqn:photon_snr}
\end{equation}
where $\alpha_i$ is the mean number of detected photons for a single measurement of spin state $\ket{i}$.
 Here, we assume $\alpha_0>\alpha_1$ and that the noise in each signal is dominated by photon shot noise, with variance $\sigma_i^2=\alpha_i$.
 The SNR is related to the dimensionless contrast between the two signals,
\begin{equation}\label{eq:Contrast}
C = \left(1 - \frac{\alpha_1}{\alpha_0}\right),
\end{equation}
such that the photon-summation SNR can be recast as:
\begin{equation}
\textrm{SNR} = \sqrt{\alpha_0}\times \frac{C}{\sqrt{2-C}}.
\label{eqn:contrast_snr}
\end{equation}
 
 This formulation clearly separates the SNR's dependence on photon collection efficiency and spin contrast.
 Note that our definition of $C$ differs from the related metric used by some authors  which we term the visibility, $V~=~(\alpha_0-\alpha_1)/(\alpha_1+\alpha_0)$.
Adding to potential confusion, the dimensionless parameter $C$ defined in the seminal work by Taylor et al. \cite{Taylor2008} is neither the contrast, nor the visibility, but is rather the inverse of the spin-readout noise, discussed in Section \ref{sec:SpinReadoutNoise}.
 For the case of NV centers, it is natural to define the contrast as in Equation~(\ref{eq:Contrast}) since $\alpha_0$ is related to the optically pumped initial spin state and often appears in defining the normalized PL, $S/\alpha_0$.
 For an NV center in bulk diamond, typically $C\approx0.3$ using the traditional PL-based readout approach.
 In the limit of perfect contrast ($C=1$), \mbox{the photon-summation} SNR is limited by shot noise alone.

\subsection{Thresholding}
 If many photons are detected during a single measurement cycle, the photon summation technique becomes less efficient than assigning a discrete outcome based on a threshold condition \cite{DAnjou2014}.
 In this scenario, the signal is modeled by the sum of two photon probability distributions (typically Poissonian or Gaussian) with different means.
 A threshold value is selected to distinguish between the two distributions, resulting in a binomial random variable specifying the outcome zero or one.
 For example, suppose the $\ket{0}$ state generates a detected number of photons that exceeds the threshold (yielding binary $S=1$) with probability $p_{0|0}$, whereas $\ket{1}$ generates a detection event that exceeds the threshold with probability $p_{0|1}$.
 Here, $p_{0|0}$ is the true positive rate, implying a false negative rate $\epsilon_0 = 1-p_{0|0}$, whereas $\epsilon_1=p_{0|1}$ is the false positive rate.
 The readout fidelity, a measure of the confidence in a given measurement outcome, is defined in terms of these two error rates as \cite{DAnjou2014,DAnjou2016}:

\begin{equation}
\mathcal{F} = 1 - \frac{1}{2}\left(\epsilon_0 + \epsilon_1\right).\label{eqn:fidelity}
\end{equation}
The fidelity takes values between 50$\%$ and 100$\%$, assuming an optimal threshold condition has been selected.


 The binomial nature of thresholded readout facilitates the direct evaluation of the signal mean and variance for an initial spin state $\ket{i}$,
\begin{equation}
\braket{ S_i } = p_{0|i}
\label{eqn:threshold_mean}
\end{equation}
\begin{equation}
\sigma^2_i = p_{0|i}(1-p_{0|i}),
\label{eqn:threshold_var}
\end{equation}
from which we can calculate the corresponding differential SNR directly from Equation~(\ref{eqn:diff_snr}):

\begin{equation}
\textrm{SNR} = \frac{p_{0|0}- p_{0|1}}{\sqrt{p_{0|0}(1-p_{0|0}) + p_{0|1}(1-p_{0|1})}}.
\label{eqn:binomial_snr}
\end{equation}
 Assuming symmetric error probabilities, $\epsilon_0=\epsilon_1$, Equation~(\ref{eqn:binomial_snr}) takes the simplified form: 
 \begin{equation}
 \textrm{SNR} = \frac{2\mathcal{F} - 1}{\sqrt{2\mathcal{F}(1-\mathcal{F})}}.
 \end{equation}
 This formulation provides a standard criterion, sometimes quoted in the literature, for determining whether a quantum state readout is single-shot; a readout fidelity $\mathcal{F}>79\%$ corresponds to an \mbox{SNR $>$ 1}.

Oftentimes, the measured value of $\mathcal{F}$ is less than would be predicted from the ideal signal \mbox{SNR \cite{Robledo2011,Magesan2015,Harty2014}}.
 This discrepancy stems from backaction (unwanted state changes during the measurement) and also potentially from improper state initialization.
 We will discuss these issues below in the context of different readout techniques.

\subsection{Spin-Readout Noise\label{sec:SpinReadoutNoise}}

 In a quantum sensor, the environmental state is mapped onto the qubit state such that the information is contained in a population difference, resulting in a stochastic signal whose mean is given by:

\begin{equation}
\langle S \rangle = \cos^2\left(\frac{\theta}{2}\right)\langle S_0 \rangle + \sin^2\left(\frac{\theta}{2}\right)\langle S_1 \rangle.
\label{eqn:generic_signal}
\end{equation}
Here, the angle $\theta$ depends on some external field (resulting, for example, from free evolution under an external magnetic field, $B$, such that $\theta \propto B$).
 The minimum resolvable angular shift, $\delta\theta$, corresponds to the situation when the change in signal exceeds the noise, $\sigma_S$. Mathematically,

\begin{equation}
\delta\theta = \frac{\sigma_S}{\left|\frac{\partial\langle S\rangle}{\partial\theta}\right|}. 
\label{eqn:angle_deviation}
\end{equation}
For an ideal measurement, $\braket{S_0}=0$, $\braket{S_1}=1$, and $\sigma_0=\sigma_1=0$.
However, the ideal measurement still exhibits noise due to the stochastic projection of qubit states.
This projection noise is the basis for the standard quantum limit (SQL) for detecting angular shifts in a single measurement.
Since projection is a binomial process, the variance of the signal depends on $\theta$, similarly to the case of Equation~(\ref{eqn:threshold_var}) for thresholded measurements:
\begin{equation}
\sigma_{\textrm{SQL}} = \sqrt{p_0(\theta)[1-p_0(\theta)]} = \frac{1}{2}\sin(\theta).
\end{equation}
Since the change in signal varies identically,
\begin{equation}
\frac{\partial\langle S_{\textrm{SQL}}\rangle}{\partial\theta}=\frac{1}{2}\sin(\theta),
\end{equation}
the SQL for a single-shot measurement is a constant angle given by $\delta\theta_{\textrm{SQL}} \equiv \theta_0 =1$ radian.

Given this fundamental limit, it is illustrative to define a parameter that quantifies the effect of realistic, imperfect measurements. The spin-readout noise, 
\begin{equation}
\sigma_R \equiv \frac{\sigma_S}{\left|\frac{\partial\langle S\rangle}{\partial\theta}\right|\theta_0},
\end{equation}
is a dimensionless quantity $\geq$1, where a value $\sigma_R=1$ signifies a measurement at the SQL \cite{Taylor2008,Shields2015}.
\mbox{The minimum} experimentally-resolvable angular shift is then given by:

\begin{equation}
\delta\theta = \theta_0\sigma_R.
\end{equation}
This formulation explicitly separates the resolution limit into two categories: the quantum mechanical noise ($\theta_0$) and experimental noise ($\sigma_R$).
 A related metric, also called the readout fidelity by some authors \cite{Taylor2008,Lovchinsky2016}, is simply the inverse, $\sigma_R^{-1}$.
 This definition of readout fidelity spans the range $(0, 1]$, where unity indicates an ideal measurement, and it differs fundamentally from the traditional definition of quantum readout fidelity (Equation~\ref{eqn:fidelity}).
 We use the traditional definition for $\mathcal{F}$ in the remainder of this work.

\subsection{Averaging}
The preceding discussion concerns single-shot readout of individual qubits. In many cases, it is advantageous to repeat the measurement (including, usually, a full experimental cycle of initialization and coherent evolution) many times in order to identify small signals.
 This is especially true when the single-shot SNR is well-below unity.
 Assuming independent trials, the SNR formulation provides a simple means for calculating the time-averaged SNR, namely,

\begin{equation}
\langle\textrm{SNR}\rangle = \sqrt{N}\times\textrm{SNR},
\label{eqn:time_avg_snr}
\end{equation}
where $\langle\rangle$ signifies the time-average and $N$ is the number of measurements.
 The parameter $N$ can account for measurements averaged in space (for ensembles of identical qubits) or time (for repeated measurements).
 In the remainder of this review, we consider especially the case of time-averaging, where $N$ is related to the total integration time, and Equation~(\ref{eqn:time_avg_snr}) allows for the direct comparison of different measurement techniques while accounting for the overhead from varying measurement durations.
 Especially for sensing applications, it bears remembering that qubit ensembles offer an additional improvement that scales with the square root of the ensemble size.

\subsection{Sensitivity}
 Sensors generally aim to acquire as much information as possible about an environmental state before it changes.
 Accordingly, we must quantify the tradeoff between signal amplitude and measurement bandwidth.
 Usually, signals are averaged over many experimental cycles, and it is useful to define the field sensitivity,

\begin{equation}
\eta = f(\theta_0)\sigma_R\sqrt{\tau},
\label{eqn:sensitivity}
\end{equation}
where the function $f(\theta_0)$ relates the SQL to a particular field amplitude, and $\tau$ is the time it takes to perform a single measurement cycle, including initialization, operation, and readout.
 The sensitivity has dimensions of $[\mathrm{field\,amplitude}]\cdot\si{\hertz}^{-1/2}$, and the minimum resolvable field can be estimated by dividing $\eta$ by the square root of total integration time.
 Barring additional noise sources or instability in the field to be measured, arbitrarily low fields can be resolved by integrating for longer times.

 Two common sensing applications are the detection of dc and ac magnetic fields \cite{Taylor2008,Degen2017}.
For the case of dc magnetic fields, the field amplitude is mapped onto a quantum phase difference using a Ramsey sequence, with a corresponding SQL given by:
 \begin{equation}
 f_{B_\mathrm{dc}}(\theta_0) = \frac{\hbar}{g\mu_BT^*_2}\theta_0,
 \label{eqn:dc_sensitivity}
 \end{equation}
 where $g$ is the gyromagnetic ratio, $\mu_B$ is the Bohr magneton, and $T_2^*$ is the inhomogeneous spin dephasing time.
 Dropping the factor $\theta_0=1$, the corresponding sensitivity is:
 
\begin{equation}
 \eta_{B_\mathrm{dc}} = \frac{\hbar}{g\mu_B}\sqrt{\frac{T_2^* + t_I + t_R}{(T_2^*)^2}}\sigma_R,
 \end{equation}
 where $t_I+t_R$ is the time required to initialize and read out the spin state, which will be referred to as measurement overhead in this review.
 Similarly, oscillating magnetic fields are detected by implementing a Hahn echo or dynamical decoupling sequence to accumulate phase. In this case, \mbox{the ac} field resolution is:
 
 \begin{equation}
 f_{B_\mathrm{ac}}(\theta_0) = \frac{\pi\hbar}{2g\mu_BT_2}\theta_0,
 \end{equation}
where $T_2$ is the homogeneous spin dephasing time, and the corresponding sensitivity is:
\begin{equation}
 \eta_{B_\mathrm{ac}} = \frac{\pi\hbar}{2g\mu_B}\sqrt{\frac{T_2 + t_I + t_R}{(T_2)^{2}}}\sigma_R.
 \end{equation}

 In general, both $\sigma_R$ and $\eta$ depend on the average value of $\theta$ at which the measurement is performed. In most cases, however, the optimum conditions for sensing are very close to $\theta=\pi/2$. Making this assumption, we derive the following analytic expressions for the spin-readout noise for the cases of photon summation,

\begin{equation}\label{eqn:photon_readout_noise}
\sigma_R^{\textrm{Photon}} = \sqrt{1 + 2\frac{\alpha_0 + \alpha_1}{(\alpha_0-\alpha_1)^2}},
\end{equation}
and for thresholding,
\begin{equation}\label{eqn:threshold_readout_noise}
\small
\sigma_R^{\textrm{Threshold}} = \sqrt{1 + 2\frac{p_{0|0}\left(1-p_{0|0}\right) + p_{0|1}\left(1-p_{0|1}\right)}{\left(p_{0|0}-p_{0|1}\right)^2}}.
\end{equation}
  Derivations are included in Appendix~\ref{appendix:sigmaR_calculations}.
 In both cases, the spin-readout noise is directly related to the differential SNR, following the general expression:

\begin{equation}
\sigma_R = \sqrt{1 + \frac{2}{\textrm{SNR}^2}}.
\label{eqn:projection_noise}
\end{equation}
 The combination of Equation~(\ref{eqn:projection_noise}) with Equation~(\ref{eqn:sensitivity})
provides a general approach to calculate the sensitivity for all spin-readout techniques covered in this review, while also accounting for variable readout durations where the SNR further becomes a function of $\tau$ (discussed in Section~\ref{sec:scc}).

\subsection{Summary}
 Particular applications benefit from different aspects of the spin-readout metrics described in the previous subsections.
 For example, quantum algorithms generally demand single-shot readout with small error probabilities. 
 Therefore, readout fidelity is the most informative choice.
 Magnetometry and sensing applications, on the other hand, usually rely on time-averaging and are inherently subject to the standard quantum limit; in this case, spin-readout noise is the most illuminating metric.
 Each of these metrics can be related to the SNR, which serves as a useful basis of comparison across multiple techniques.
 Table~\ref{table:metrics} summarizes the three metrics discussed in this section and their relation to SNR.

 In somel situations, a critical experimental design consideration is whether to use photon summation or thresholding.
 To decide, we can compare the thresholding SNR (Equation~(\ref{eqn:binomial_snr})) to the photon summation SNR (Equation~(\ref{eqn:photon_snr})) and choose the higher value.
 Typically, thresholding becomes more efficient when one of the spin states produces $>$1 photon in a measurement and the contrast exceeds 50$\%$.
 We hope that the connections between these metrics and various measurement techniques described in the following sections will aid in selecting the optimal approach for \mbox{future applications}.


\begin{table*}[t]
\centering
\ra{1.75}
\begin{ruledtabular}
\begin{tabular}{@{}lcl@{}}

\textbf{Metric} & \textbf{Relation to SNR} & \textbf{Use Case} \\ \hline
Contrast, $C$, \& Count rate, $\alpha_0$ & $\textrm{SNR}=\sqrt{\alpha_0}\frac{C}{\sqrt{2-C}}$ & traditional PL readout \\
Spin-readout noise, $\sigma_R$ & $\textrm{SNR}=\sqrt{\frac{2}{\sigma_R^2-1}}$ & magnetometry \\
Fidelity, $\mathcal{F}$ & $\textrm{SNR}=\frac{p_{0|0}- p_{0|1}}{\sqrt{p_{0|0}(1-p_{0|0}) + p_{0|1}(1-p_{0|1})}}$ & quantum algorithms, large signals \\
Repeats for $\braket{\textrm{SNR}}=1$ & $N=\left(\frac{1}{\textrm{SNR}}\right)^2$ & magnetometry, general experiments \\

\end{tabular}
\end{ruledtabular}
\caption{Compilation of spin-readout metrics, their formal relation to differential SNR, and common use cases. \label{table:metrics}}
\end{table*}

\section{Traditional Spin Readout \label{sec:traditional_readout}}

 The NV center's intrinsic, spin-dependent PL facilitated the first room-temperature quantum control experiments with single spins \cite{Gruber1997,Jelezko2004}.
 Simply by counting the PL photons emitted in the first $\sim$\SI{300}{\nano\second} of optical illumination and averaging over many cycles, the NV center's ground-state spin projection can be inferred.
 This technique, here called traditional PL readout, is still widely used in research and applications due to its simple experimental implementation.
 This section outlines the physical mechanisms that underlie traditional PL readout, as well as some of the \mbox{technique's limitations}.

 The negatively-charged NV center is a point defect with $C_{3v}$ symmetry that exhibits isolated electronic states deep within the diamond's band gap including a paramagnetic triplet ground \mbox{state \cite{Doherty2013}}.
 The $C_{3v}$ symmetry axis points along any of the $\braket{111}$ crystallographic axes, connecting the substitutional nitrogen and adjacent vacancy. 
 The broken inversion symmetry leads to a zero-field energy splitting between the ground state's $m_s=0$ and $m_s=\pm1$ spin sub-levels (\SI{2.87}{\giga\hertz} at room temperature, with energies here and throughout given in frequency units), and a dc magnetic field applied along the defect's symmetry axis further splits the $m_s=\pm1$ levels such that individual transitions can be addressed using spin resonance techniques.
 This yields the commonly-used qubit manifolds, encompassing the $m_s=0$ state and one of the $m_s=\pm1$ projections.
 Diamond's low nuclear-spin density and weak spin-phonon coupling allow for the NV center's spin coherence to reach milliseconds at room temperature \cite{Balasubramanian2009}.
 The long spin coherence times allow for the detection of weak magnetic fields \cite{Taylor2008}, including those associated with proximal nuclear \cite{Childress2006} and electron \cite{Dolde2013} spins, enabling the realization of multi-qubit quantum registers \cite{Neumann2008,Taminiau2014}.

\begin{figure*}[t]
\includegraphics[scale=0.9]{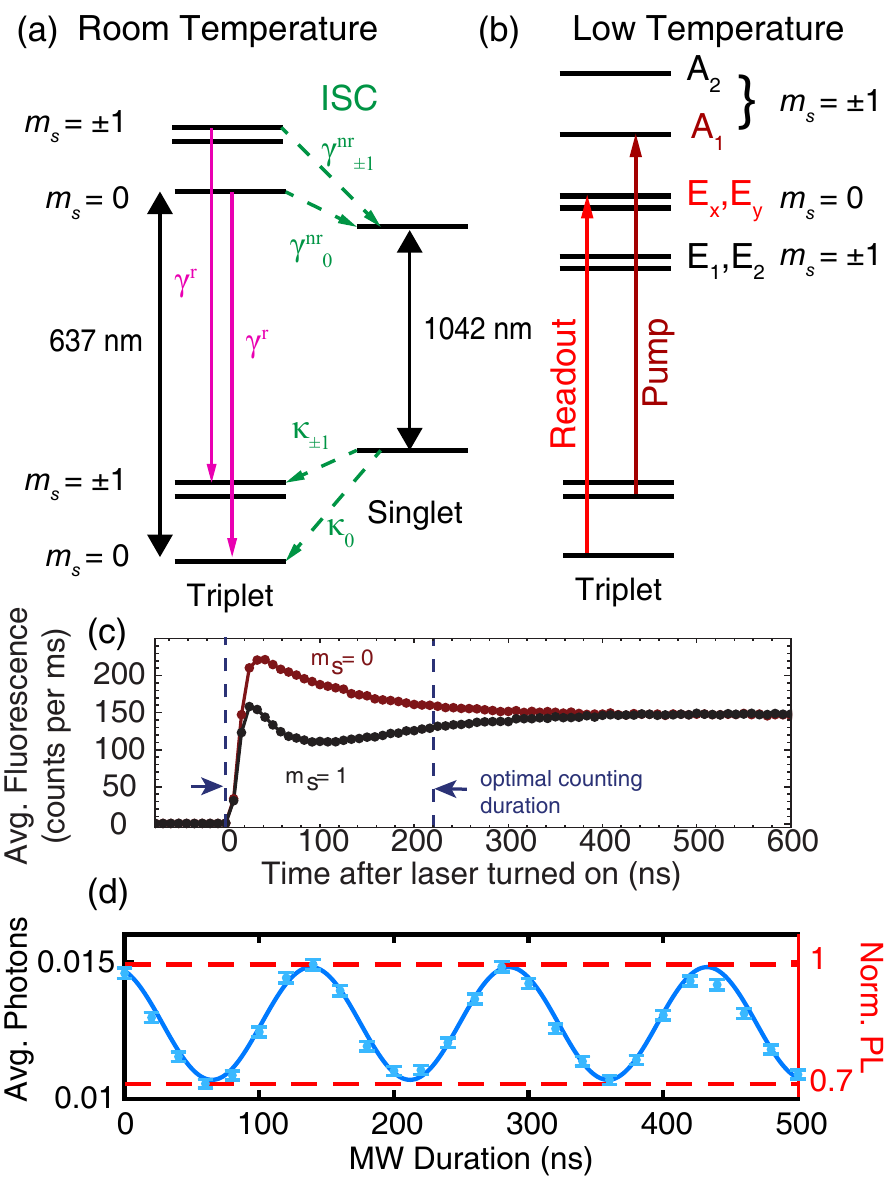}
\caption{\label{fig:level_diagram} The diamond NV center. (\textbf{a}) Room temperature electronic structure. Solid lines indicate radiative transitions (with corresponding rate $\gamma^\mathrm{r}$), and dashed lines represent nonradiative intersystem crossing (ISC) transitions (with rates $\gamma^\mathrm{r}_i$ and $\kappa_i$ for the excited and ground-state spin projection $i$, respectively). Solid black arrows represent the zero phonon lines of the triplet and singlet manifolds. (\textbf{b}) Low temperature electronic structure of the nitrogen-vacancy (NV) center triplet manifold. Individual transitions used for spin pumping and resonant readout are indicated. (\textbf{c}) Room temperature transient fluorescence response for the spin states $m_s=0,1$ produced by \SI{532}{\nano\meter} illumination. The optimal counting duration is indicated by the dashed vertical lines. Reprinted with permission from \cite{Gupta2016}, Optical Society of America. (\textbf{d}) Rabi nutations of the ground-state spin at room temperature, measured using traditional PL readout for an NV center beneath a planar diamond surface, with an NA = 0.9 objective. The left and right axes plot the average detected photons per measurement and normalized PL, respectively. The solid curve is a fit to the data.}
\end{figure*}

Under visible illumination (typically \SI{532}{\nano\meter}), the NV center emits PL in its $\approx$650--\SI{750}{\nano\meter} phonon sideband (PSB) whose intensity depends on the ground-state spin projection.
 Physically, spin-dependent PL arises from spin-orbit interactions within the intersystem crossing (ISC) that couples the triplet and singlet manifolds \cite{Oort1988, Goldman2015a}.
 As shown in Figure~\ref{fig:level_diagram}a, the excited-state triplet levels can undergo radiative transitions back to the ground state or nonradiatively decay into the meta-stable singlet manifold.
 The total decay rate of the excited state spin projection $\ket{i}$ is given by the sum of these two rates, namely:
 \begin{equation}
 \gamma_i = \gamma^{\textrm{r}} + \gamma_i^{\textrm{nr}}.
 \label{eqn:gamma_total}
 \end{equation}
The radiative rate, $\gamma^\mathrm{r}$, is essentially spin independent, whereas the nonradiative rates, $\gamma_i^\mathrm{nr}$, depend strongly on the spin projection due to the spin-dependent ISC.
 Recent studies concluded that $\gamma_{\pm1}^{\textrm{nr}} \approx 10\gamma_0^{\textrm{nr}}$ \cite{Goldman2015a,Gupta2016,Hopper2016}.
This difference produces a transient response to illumination that is drastically different depending on the initial projection of the ground-state spin.
 
 Assuming the NV center is illuminated with an optical excitation rate similar to $\gamma^\mathrm{r}$ (i.e., close to optical saturation, which is generally ideal for traditional PL readout), a spin population initially in $m_s=\pm1$ is shelved into the singlet manifold within only a few optical cycles of the triplet states, whereas a population in $m_s=0$ continues to cycle and produce PL.
 This spin-dependent PL contrast is the essence of traditional readout.
 The contrast is short-lived, however; it vanishes after about \SI{300}{\nano\second} as the singlet population decays back to the triplet ground-state \cite{Acosta2010}, and the system reaches a steady state (Figure~\ref{fig:level_diagram}c).
 Taking into account the spin selectivity of both the triplet-to-singlet and singlet-to-triplet ISC (the latter is less spin selective than the former), the resulting ground-state spin population after the illumination is switched off is $\approx80\%$ polarized into the $m_s=0$ sub-level \cite{Waldherr2011a, Doherty2013,robledo2011spin}.
 This optically pumped pseudo-pure state generally serves as the initialized $\ket{0}$ state for subsequent quantum experiments, while one of the $m_s=\pm1$ state serves as the $\ket{1}$ state.
 
 Figure~\ref{fig:level_diagram}d shows a typical example of room-temperature Rabi nutations for a single NV center in bulk diamond, with the data plotted in terms of both the average number of photons detected per shot and the corresponding normalized PL.
 The spin contrast is $C$ = $30\%$, and the confocal setup collects \mbox{$\alpha_0$ = 0.015} photons on average from the $\ket{0}$ spin state, using an NA = 0.9 air objective to image an NV center $\approx$ \SI{4}{\micro\meter} beneath a planar diamond surface with a saturated count rate of \SI{50}{\kilo\counts\per\second} under continuous-wave
 \SI{532}{\nano\meter} illumination.
 Using Equation~(\ref{eqn:photon_snr}), the corresponding single-shot SNR is 0.03, and Equation~(\ref{eqn:time_avg_snr}) implies that more than $10^5$ repeats are required to achieve $\langle\mathrm{SNR}\rangle=10$.
 Each point in Figure~\ref{fig:level_diagram}d consists of $4\times10^5$ repeats.
 In many applications, such averaging places severe limitations on performance and efficiency.
 In the remaining sections, we compare several alternative readout techniques to this standard, accounting for experimental variations in collection efficiency where possible.

%

\section{Maximizing Photon Collection Efficiency \label{sec:collection_efficiency}}
 The NV's optical addressability in a solid-state host material provides both technological opportunities and formidable engineering challenges.
 Due to the high refractive index of diamond ($n\approx2.4$), total internal reflection at diamond-air interfaces severely limits the collection efficiency; even assuming an air objective with NA = 0.95, a maximum fraction of only $4\%$ of emitted photons can be extracted through a planar (100)-oriented surface \cite{PLAKHOTNIK199583}.
 Since the spin-readout noise is dominated by the Poisson statistics associated with counting photons, collection efficiency improvements that increase $\alpha_0$ boost the single-shot SNR according to $\sqrt{\alpha_0}$ (Equation~(\ref{eqn:contrast_snr})) and reduce the averaging requirements according to $N\propto 1/\alpha_0$.
 This section considers strategies for improving the collection efficiency of NV centers within bulk diamond.

\subsection{Crystal Alignment\label{sec:CrystalAlignment}}
 The NV center's optical dipoles are oriented perpendicularly to the symmetry axis connecting the nitrogen atom to the vacancy.
 Since the symmetry axis points along a crystalline $\langle111\rangle$ direction, aligning the optical axis perpendicularly to the corresponding \{111\} face maximizes optical absorption and emission.
 However, the (100) orientation of most commercially available synthetic diamonds misaligns the NV's symmetry axis by \SI{55}{\degree} from the optical axis.
 Using a 100x, NA = 0.9 air objective, Jamali et al. \cite{Jamali2014} showed that proper alignment of the dipole and optical axes results in a $65\%$ increase in collected photons, corresponding to an SNR increase of $\approx$1.3.
 Although the production of (111)-faced diamonds is traditionally a laborious and expensive process, recent developments of laser-nucleated-cleaving techniques \cite{Parks2018} provide an attractive alternative.
 In this technique, a series of laser pulses is used to nucleate and propagate cleaves along desired (111) planes within a standard (100)-faced diamond, resulting in large, flat, (111)-faced plates even without any polishing.
 Ideally-oriented NVs can then be produced using standard electron-irradiation or nitrogen implantation techniques, followed by annealing.
 Furthermore, recent studies have shown that growth of diamond along $\langle 111\rangle$ directions can yield deterministically-oriented NVs with the optimum alignment perpendicular to the (111) surface \cite{Miyazaki2014,Lesik2014,Michl2014,Fukui2014,Ozawa17}.

\subsection{Photonic Structures\label{sec:PhotonicStructures}}

 \begin{figure*}[t]
\includegraphics[scale=0.9]{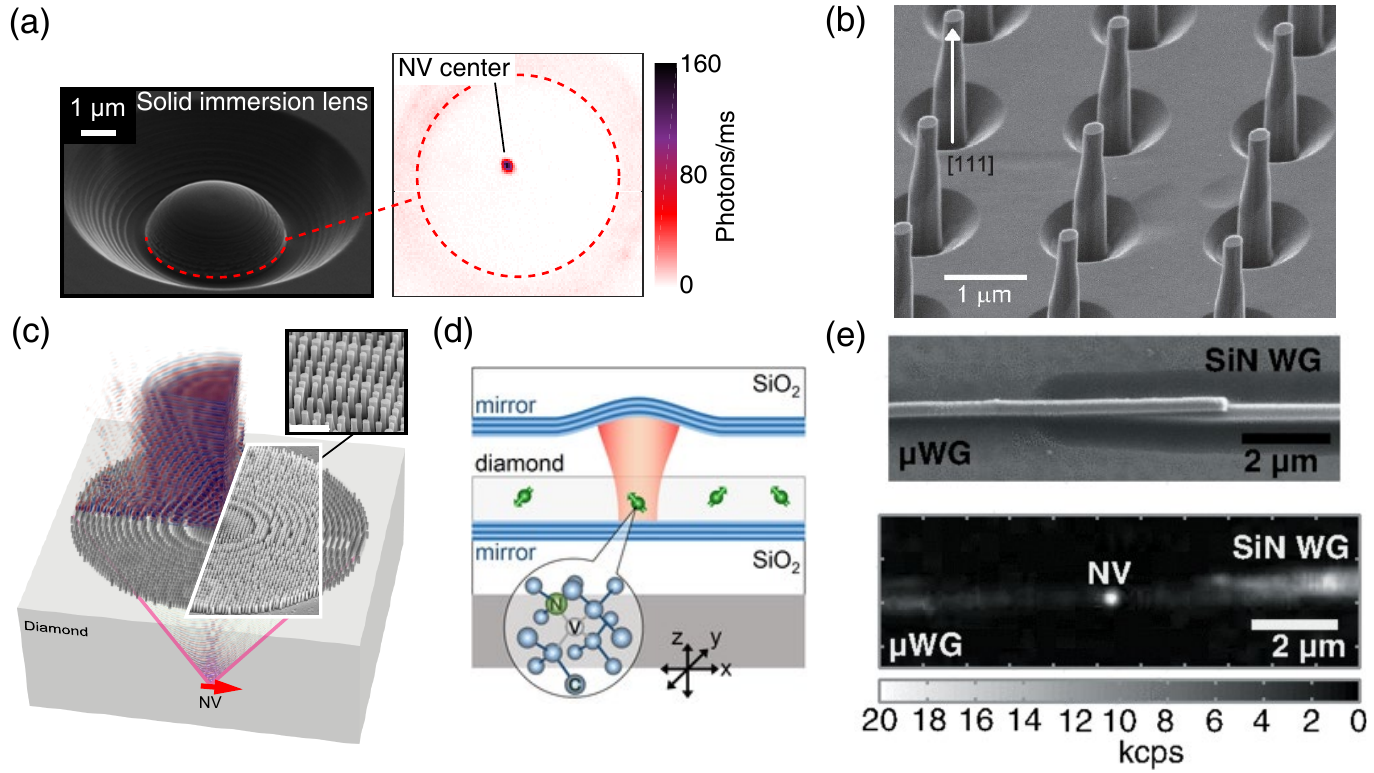}
\caption{\label{fig:photon_collection} {Photonic devices for improving collection efficiency} (\textbf{a}) Scanning electron micrograph (SEM) of a solid immersion lens fabricated around an NV center. Inset: confocal PL image. \mbox{(\textbf{b}) Diamond} nanopillar array fabricated on a [111]-oriented diamond. Source: Neu et al. \cite{Neu2014}. (\textbf{c}) Metalens fabricated above an NV center to act as an immersion objective. Inset: SEM of nanopillar metalens elements. Source: Grote et al. \cite{Grote2017}. (\textbf{d}) Schematic of diamond membrane embedded in an open micro-cavity. Source: Riedel et al. \cite{Riedel2017}. (\textbf{e}) SEM of a hybrid diamond/silicon-nitride waveguide and a PL map of an NV center within the diamond waveguide. Source: Mouradian et al. \cite{Mouradian2015}.}
\end{figure*}

 Advances in nanofabrication and photonic design have produced several top-down fabrication solutions to circumventing the diamond-air refractive index mismatch.
 The solid immersion lens (SIL), consisting of a hemisphere etched around an NV center (Figure~\ref{fig:photon_collection}a), overcomes total internal reflection such that only Fresnel reflection contributes to losses \cite{Hadden2010,Marseglia2011,Jamali2014}, and the latter can be further reduced using antireflective coatings.
 When used together with proper orientation of the diamond crystal (Section~\ref{sec:CrystalAlignment}), a SIL can increase the saturation count rate to over \SI{1}{\mega\counts\per\second} \cite{Jamali2014,Robledo2011}, resulting in an overall SNR improvement of a factor of five as compared to an NV in a (100)-oriented planar sample.
 Recently, a metalens constructed from nanopillars etched on the diamond surface was used to image an NV center \cite{Grote2017}.
 In contrast to the SIL, the metalens design collimates the emitted light (Figure~\ref{fig:photon_collection}c), removing the need for a free-space objective and making it a promising approach towards coupling NV centers directly with optical fiber.

 An alternative method involves embedding an NV center directly within a diamond pillar or nanowire \cite{Babinec2010}.
 The waveguiding effect of the nanopillar directs the emission normal to the diamond surface.
 An example nanopillar on a [111]-oriented diamond substrate is depicted in Figure~\ref{fig:photon_collection}b \cite{Neu2014}.
 The photons can be collected using an air or oil-immersion objective, with count rates exceeding \SI{1}{\mega\counts\per\second} \cite{Momenzadeh2015}.
 The nanopillar design has been utilized in improving the sensitivity of scanning magnetometers \cite{Maletinsky2012,Appel2016}.
 A related nanophotonic design is the nanobeam \cite{Shields2015}, which directs emission from embedded NV centers into an underlying substrate and has also yielded saturation count rates $>$\SI{1}{\mega\counts\per\second}.
 \mbox{In each} of these cases, the high collection efficiency comes at the cost of fabrication complexity, often with the requirement for precise NV alignment relative to the photonic structure.
 \mbox{In the case} of nanopillars, nanobeams, and other nanophotonic structures that incorporate NV centers close to etched surfaces, detrimental effects from charge and spin noise at the diamond surface further impede performance by reducing the NV center's optical and spin coherence properties.

 \begin{figure*}[t]
\includegraphics[scale=0.8]{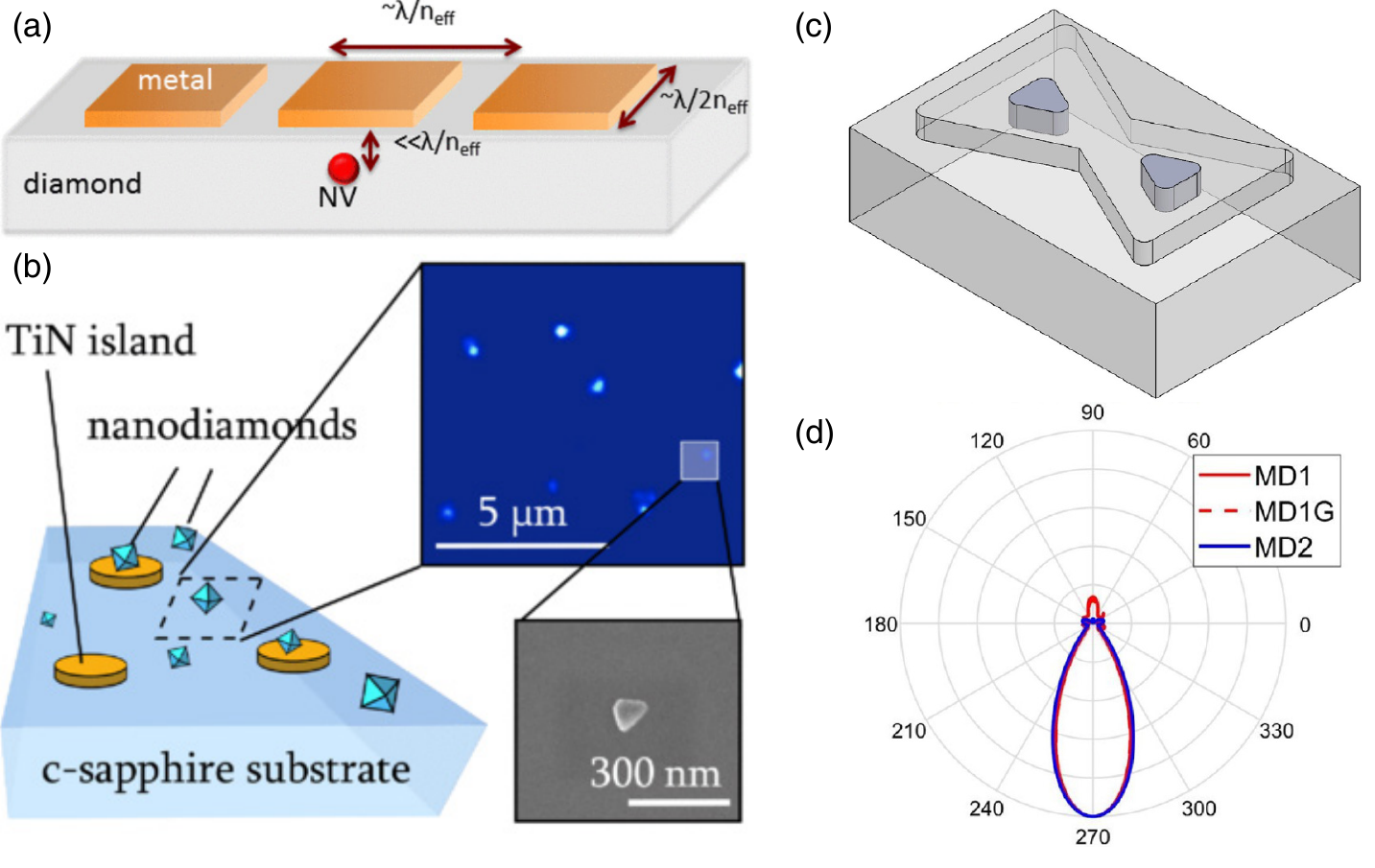}
\caption{\label{fig:radiative_lifetime} {Radiative lifetime engineering with plasmonic devices.}
Recent examples of plasmonic device geometries include: (\textbf{a}) a shallow NV center situated below an optical plasmonic antenna; \mbox{(\textbf{b}) nanodiamonds} containing NV ensembles deposited over TiN plasmonic resonators; and (\textbf{c}) a hybrid dielectric-metal hourglass structure designed to couple to a shallow NV. Panel (\textbf{d}) shows the highly directional angular emission distribution that results from hybrid hourglass plasmonic devices (the device shown in ({c}) is labeled MD2). ``MD'' stands for metal-dielectric. See~\cite{Karamlou2018} for details on the design variations in ({d}). Panel ({a}) is reprinted with permission from \cite{Wolf2015}. Copyright 2015 by the American Physical Society. Panel ({b}) is reprinted with permission from \cite{Bogdanov2017}. Copyright 2017 by the American Physical Society. Panels ({c},{d}) are from Karamlou et al. \cite{Karamlou2018}.}
\end{figure*}

\subsection{Waveguides and Cavities}
 Integrated single-mode diamond waveguides \cite{Hausmann2012,Gould2016} have enabled on-chip optical and spin control of single NVs \cite{Mouradian2015} with saturation count rates approaching \SI{1}{\mega\counts\per\second} (Figure~\ref{fig:photon_collection}e).
 Diamond waveguides can be fabricated using a variety of techniques, with the most common being a diamond-on-insulator approach in which a thin diamond membrane is placed on a lower-refractive-index substrate and patterned using top-down lithography and dry etching \cite{Schroder:16}.
 Micro-ring resonators \cite{faraon2011resonant} and photonic crystal cavities have been realized in a similar fashion \cite{Hausmann2013,Faraon2012}, both of which exhibit Purcell enhancements (discussed in Section~\ref{sec:radiative_lifetime}) due to the high quality factor of the dielectric cavities.

 Due to the relatively small size required for single-mode operation ($<$300 nm), integrated photonic devices suffer from the same challenges due to fabrication damage and enhanced surface noise as the nanopillar and nanobeam structures discussed in Section~\ref{sec:PhotonicStructures}.
 Furthermore, technical issues associated with submicron diamond membranes (e.g., enhanced strain, nonparallel surfaces and laborious fabrication requirements) have impeded the widespread adoption of these approaches.
 New designs and fabrication approaches that allow waveguides and cavities to be created directly from bulk diamond crystals \cite{Grote2016, mouradian2017tunable} potentially offer a way forward, although the control of surface noise that causes deteriorated optical linewidths in nanophotonic structures remains a formidable challenge.
 One approach to avoiding these sources of noise is to use NVs embedded within diamond membranes of micron-scale thickness, which can be aligned within high-finesse fiber-based cavities, albeit with larger mode volumes (Figure~\ref{fig:photon_collection}d) \cite{Johnson2015,Bogdanovic2017,Riedel2017}.

\subsection{Summary}
 In addition to traditional PL spin readout, every technique described in this review gains performance improvements by increasing the photon collection efficiency.
 However, constructing optimized structures remains a barrier due to the difficulties associated with nanofabrication of diamond.
 Detrimental surface effects on the spin and optical coherence properties of shallow NV centers need to be mitigated.
 Ultimately, in the limit of near-unity collection efficiencies, detector dead times will become a limiting factor to achieving single-shot fidelities, and the use of multiple detectors may be necessary.
 Overcoming these design, fabrication, materials, and measurement challenges will play a critical role in the development of NV-based quantum devices.

\pagebreak
\section{Radiative Lifetime Engineering \label{sec:radiative_lifetime}}

A potential alternative approach to increasing the number of detected photons relies on nanophotonic engineering of the local density of optical states. 
 Dielectric or plasmonic structures can decrease radiative lifetime and increase the photon emission rate through the Purcell effect \cite{Purcell1946}.
 \mbox{The ability} to incorporate quantum emitters within nanophotonic devices has spurred recent efforts to investigate the limits of the Purcell effect, and large gains have been reported \cite{Schroder:16, Vahala2003}.
 The potential for radiative-lifetime engineering to improve the NV center's optical spin readout efficiency has theoretically been predicted \cite{Wolf2015}, but experimental verification is missing.
 Since the SNR depends on both the photon count rate and spin contrast (Equation~(\ref{eqn:contrast_snr})), a better understanding of the optical dynamics in the limit of high Purcell enhancement is required.
 Here, we provide an overview of current research in this area and highlight several unanswered questions.
 
 Due to their small optical mode volume, dielectric photonic crystal cavities can drastically increase the optical density of states for an embedded NV center \cite{Wang2007,Wolters2010,Faraon2012,Hausmann2013,barclay2009coherent,Lee2014}.
 The cavity not only directs the far-field emission, but also decreases the radiative lifetime by an amount known as the Purcell factor,

\begin{equation}\label{eq:Purcell}
	F_\mathrm{P} = \frac{3}{4\pi^2}\left(\frac{\lambda}{n}\right)^3\left(\frac{Q}{V}\right),
\end{equation}
where $\lambda$ is the free-space wavelength, $n$ is the refractive index, $Q$ is the quality factor, and $V$ is the mode volume.
 Equation~(\ref{eq:Purcell}) represents the ideal case, assuming a cavity mode resonant with the relevant optical transition and an optical dipole located at the position of maximum field, aligned with its polarization axis.
 In practice, NV centers can be directly embedded in photonic crystal cavities fabricated from thin diamond membranes \cite{Faraon2012,Hausmann2013,Lee2014,Riedel2017,Bogdanovic2017} or positioned close to cavities fabricated in another high-refractive-index material \cite{Wolters2010,Englund2010,barclay2009coherent}.
 The prior method generally results in higher $F_\mathrm{P}$ than the latter, due to increased spatial overlap between the NV center's optical dipole and the cavity \mbox{field \cite{Hausmann2013}}.
 Most investigations have explored how the zero-phonon-line emission around \SI{637}{\nano\meter} can be enhanced \cite{Wolters2010,Faraon2012,Hausmann2013,Riedel2017,Bogdanovic2017}, since photons in this band are ultimately required for coherent spin-photon interfaces with NV centers.
 Meanwhile, potential effects on the spin-readout SNR for NV centers coupled to photonic crystal cavities remain relatively unexplored.

 NV centers placed in close proximity to plasmonic resonators can also exhibit large Purcell \mbox{factors \cite{Kuhn2006,Akimov2007,Schietinger2009}}.
 The extreme spatial confinement of plasmons can boost $F_\mathrm{P}$ through a strong reduction of $V$ in Equation~(\ref{eq:Purcell}), even when $Q$ is generally lower for plasmonic as compared to dielectric \mbox{structures \cite{akselrod2014probing,hoang2015ultrafast}}.
 In fact, a lower $Q$ can be desirable for coupling to broadband emission in the NV center's phonon sideband.
 As for dielectric cavities, the magnitude of the Purcell enhancement also depends on the relative orientation and location of the optical dipole and the plasmonic mode; at the same time, care must be taken to avoid quenching due to nonradiative energy transfer \cite{Anger2006}.
 The optimal metal-emitter separation depends on the material and geometry; for a gold nanoparticle, the ideal separation is $\approx$\SI{5}{\nano\meter} \cite{Anger2006}, although enhancements have been observed using nanodiamonds with buffers as thick as \SI{30}{\nano\meter} \cite{Shalaginov2013}.
 Figure~\ref{fig:radiative_lifetime}a--c shows three recent examples of plasmonic devices designed to engineer the emission dynamics of NV centers in nanodiamonds or close to the surface of bulk diamond.
 Several recent studies have further considered metal-dielectric hybrid systems that optimize both directionality and radiative lifetime reduction \cite{Bulu2011,Riedel2014,Karamlou2018}.
 Computational results predict that a hybrid bow-tie structure like the one shown in Figure~\ref{fig:radiative_lifetime}c can produce a strong Purcell enhancement together with highly directional emission (Figure~\ref{fig:radiative_lifetime}d), providing an attractive alternative to all-dielectric diffractive designs.

 The question of how Purcell enhancement affects the NV center's spin-readout SNR remains unresolved.
 Theoretical studies suggest that substantial improvements in SNR are possible \cite{Babinec2012,Wolf2015}, but the simulations depend crucially on particular transition rates between excited and ground states that have not been experimentally quantified.
 The debate centers on how shortening the radiative lifetime influences the PL contrast (see Equations~(\ref{eqn:contrast_snr}) and~(\ref{eqn:gamma_total})).
 Wolf et al. \cite{Wolf2015} showed that the SNR could increase monotonically with $F_\mathrm{P}$ if the radiative transitions are fully spin-conserving (such that the overall spin-mixing rate is unaffected by the change in radiative lifetime), whereas only incremental gains in SNR are achievable if the radiative transitions introduce spin mixing that scales with $F_\mathrm{P}$.
 A related question concerns the evolution of the NV center's ground-state spin polarization under optical illumination, which has been predicted to decrease when the radiative rate is enhanced \cite{Babinec2012}.
 Recent experiments using NV ensembles within nanodiamonds coupled to plasmonic islands (\mbox{Figure~\ref{fig:radiative_lifetime}b, \cite{Bogdanov2017}}) demonstrated that the spin-dependent PL contrast, and subsequently the SNR, decreases with increasing $F_\mathrm{P}$.
 This decrease was attributed to additional nonradiative decay pathways present for NV centers in nitrogen-rich nanodiamonds, which ultimately limits the optical excitation rate \cite{Hopper2018}.
 The situation is likely to be different for NV centers in higher-purity diamond.

Nanophotonic dielectric and plasmonic structures provide many opportunities to optimize photon emission and electromagnetic coupling properties of NV centers.
As discussed further in Section~\ref{sec:discussion}, \mbox{it can} be important to consider the ability of such structures to enhance optical absorption in addition to emission.
Although the ultimate impact of radiative lifetime engineering on spin readout remains unknown, future studies into the dynamics of Purcell-enhanced NV centers could result in significant improvements to the performance of room-temperature quantum devices.

\section{Low-Temperature Resonant Readout \label{sec:low_temp_readout}}
 The NV center's triplet excited state is an orbital doublet \cite{Tamarat2008,batalov2009low}; however, at temperatures above $\approx$\SI{20}{\kelvin}, rapid phonon-assisted orbital transitions obscure the fine structure \cite{fu2009observation}, and motional narrowing leads to an effective orbital-singlet excited-state Hamiltonian \cite{fuchs2010excited} at room temperature, as shown in Figure~\ref{fig:level_diagram}a.
 At low temperatures, however, individual spin-selective zero-phonon-line transitions connecting the ground and excited states can be resonantly addressed (Figure~\ref{fig:level_diagram}b), enabling the generation of spin-photon coherence \cite{Buckley2010, togan2010quantum} and all-optical coherent control of the NV's orbital and spin dynamics \cite{Yale2013,Bassett2014}.
 Although this review focuses on room-temperature protocols, in this section, we introduce the low-temperature resonance-fluorescence readout protocol, since it offers the highest performance currently available.

\begin{figure*}[t]
\includegraphics[scale=1]{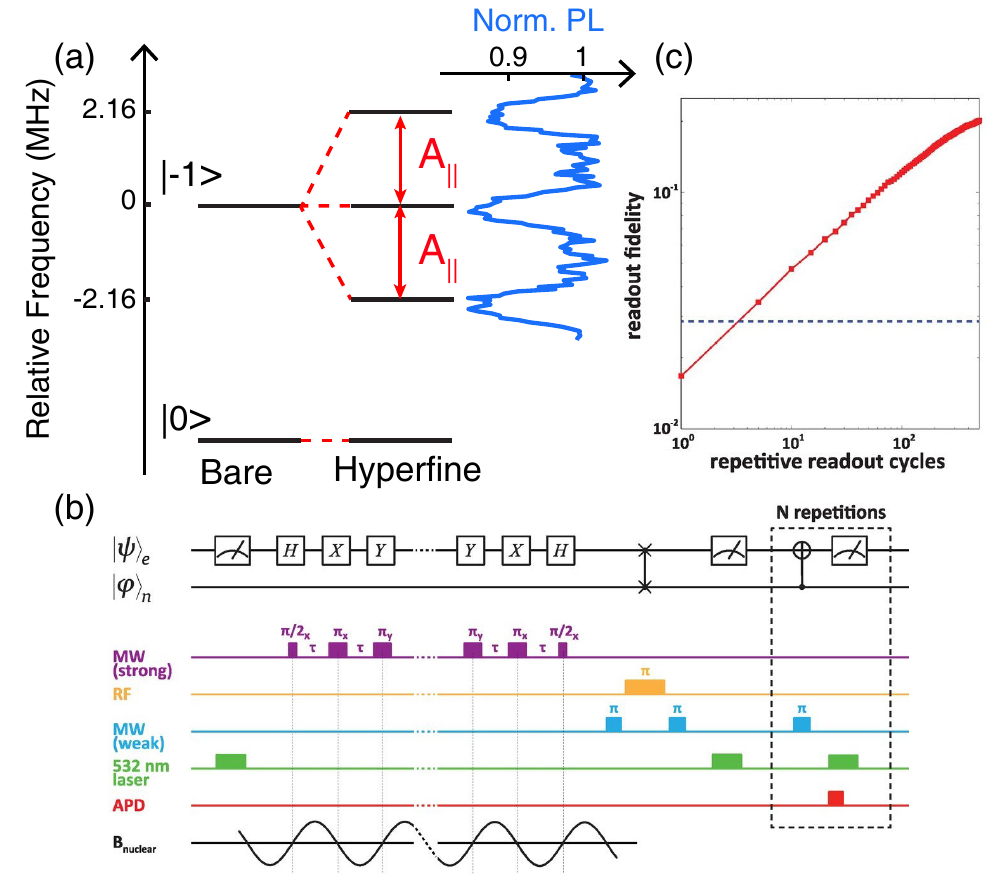}
\caption{\label{fig:nuclear_assisted} {Nuclear-assisted readout}. (\textbf{a}) Energy-level diagram showing the splitting of the $m_s=-1$ spin state into a triplet through hyperfine coupling with $^{14}$N ($A_{||}=\SI{2.16}{\mega\hertz}$). The data at the right show the normalized PL response to a pulsed electron-spin resonance measurement. (\textbf{b}) Quantum circuit and measurement timing diagram used to detect proteins on the diamond surface using a nitrogen nuclear spin as a memory for storage. (\textbf{c}) The readout fidelity, the inverse of spin readout noise (Equation~(\ref{eqn:projection_noise})), as a function of repetitive readout cycles. Panels ({b},{c}) are from \cite{Lovchinsky2016}. Reprinted with permission from AAAS.}
\end{figure*}

 In analogous fashion to protocols for resonant optical readout of trapped ions \cite{Olmschenk2007} and quantum dot molecules \cite{Vamivakas2010}, resonance fluorescence allows for single-shot readout of the NV center's electronic spin state. 
 As initially demonstrated by Robledo et al. \cite{Robledo2011}, the idea is to resonantly pump a spin-selective, spin-preserving optical cycling transition that is protected from the ISC.
 This improves both the optical contrast and the duration over which photons can be collected.
 When the external magnetic, electric and strain fields are carefully controlled \cite{bassett2011electrical}, the $m_s=0$ excited states $\ket{E_x}$ and $\ket{E_y}$ provide nearly ideal cycling transitions, producing PL photons only for the $\ket{0}$ spin state.
 Meanwhile, transitions selective for $m_s=\pm1$ spin states, such as the transition to the $\ket{A_1}$ excited state shown in Figure~\ref{fig:level_diagram}c, provide efficient optical pumping pathways to polarize the spin in $\ket{0}$ with a \mbox{99.7 $\pm$ 0.2\% probability}.
 
 In the initial demonstration \cite{Robledo2011}, resonant readout produced a measurement contrast of 89$\%$ persisting for \SI{100}{\micro\second}.
 Thresholding provides the best performance in this case; the resulting readout fidelity was 93.2$\%$, corresponding to an SNR improvement by a factor of 34 over the traditional room-temperature PL measurement shown in Figure~\ref{fig:level_diagram}d.
 Subsequent technical improvements to the resonant readout protocol such as charge stabilization, dynamical stop procedures, and better collection efficiencies have resulted in even higher readout fidelities, enabling the demonstration of quantum feedback \cite{Blok2014}, heralded entanglement \cite{Bernien2013}, loop-hole free Bell's inequality violations \cite{Hensen2015}, and quantum error correction \cite{Cramer2016}.

\section{Nuclear-Assisted Readout \label{sec:nuclear_assisted}}
 The NV center's electronic spin can interact with nearby nuclear spins.
 Prevalent nuclear species include the NV center's intrinsic nitrogen nuclear spin (with total spin $I=1$ or $\frac{1}{2}$ for the isotopes $^{14}$N and $^{15}$N, respectively) and the carbon isotope $^{13}$C (total spin $I =\frac{1}{2}$). $^{13}$C nuclei are normally present at stochastic locations proximal to the NV center due to its 1.1\% isotopic abundance.
 Nuclear spins exhibit much longer spin lifetimes than electrons \cite{Terblanche2001}, and they can be utilized as quantum \mbox{memories \cite{Childress2006,Dutt2007,Maurer2012}} and computational nodes for quantum error correction \cite{Cramer2016} and quantum \mbox{communication \cite{reiserer2016robust,kalb2017entanglement}}.
 In this section, we discuss how coupled nuclear spins can assist in improving readout of the NV center's electronic spin state \cite{Jiang2009,Steiner2010,Neumann2010}.

 The coupling between the NV center electron spin and a single nuclear spin is described by the hyperfine interaction. The hyperfine Hamiltonian can be written in the form:
 \begin{equation}\label{eq:Hhf}
 \hat{\mathcal{H}}_\textrm{hf} = A_{\parallel}\hat{S}_z\hat{I}_z + \frac{A_{\perp}}{2}\left(\hat{S}_+\hat{I}_- + \hat{S}_-\hat{I}_+\right),
 \end{equation}
 where $\hat{S}_z$ and $\hat{I}_z$ are the electron and nuclear Pauli-$z$ operators, respectively; $\hat{S}_{+/-}$ and $\hat{I}_{+/-}$ are the electron and nuclear spin raising and lowering operators, respectively; $A_{\parallel}$ is the parallel hyperfine component; and $A_{\perp}$ is the perpendicular hyperfine component.
 The magnitudes of $A_{\parallel}$ and $A_{\perp}$ depend on the two spin species, their relative orientation, and their separation.
 Physically, the parallel component represents a nuclear-spin-dependent Zeeman shift of the electron spin eigenstates, clearly observed as a splitting in the electron spin resonance spectrum, as shown in Figure~\ref{fig:nuclear_assisted}a for the case of an intrinsic $^{14}$N nuclear spin triplet with $A_\parallel=\SI{2.16}{\mega\hertz}$.
 The split resonances will be resolved as long as the hyperfine strength $A_\parallel$ exceeds the electron-spin dephasing rate, $1/T_2^\ast$.
 Such a spectrum allows for the application of nuclear-spin-selective C$_\mathrm{n}$NOT$_\mathrm{e}$ quantum gates on the electron spin, and likewise electron-spin-selective C$_\mathrm{e}$NOT$_\mathrm{n}$ gates on the nuclear spin using appropriate radio-frequency \mbox{driving fields}.

 The perpendicular component describes flip-flop interactions that mix states with \mbox{$\Delta m_s=-\Delta m_i=\pm1$}, causing unwanted electron and nuclear spin flips.
 For weakly-coupled nuclei under most conditions, flip-flop interactions are suppressed by the large zero-field splitting between electron-spin sub-levels in the NV center's ground state, and the second term in Equation~(\ref{eq:Hhf}) can be neglected; this is the so-called secular approximation.
 However, the nonsecular terms are not negligible for strongly-coupled $^{13}$C nuclei close to the defect \cite{Childress2006}, and similarly, the $A_\perp$ coupling to intrinsic $^{14}N$ and $^{15}N$ spins is substantially larger in the NV center's excited state than in its ground state due to increased overlap with the excited-state electronic orbitals.

 The basic idea of nuclear-assisted readout for NV centers, as first demonstrated by \mbox{Jiang et al. \cite{Jiang2009}}, is to harness the long spin lifetime for nuclei and the ability to correlate the electron and nuclear spin states using C$_\mathrm{n}$NOT$_\mathrm{e}$ gates, such that the PL signal from many successive readout cycles can be accumulated to amplify the SNR.
 In preparation for measurement, the electron spin state to be measured is mapped onto the nucleus using a series of C$_\mathrm{n}$NOT$_\mathrm{e}$ and C$_\mathrm{e}$NOT$_\mathrm{n}$ gates (Figure~\ref{fig:nuclear_assisted}b).
 \mbox{The readout} then consists of the repeated application of C$_\mathrm{n}$NOT$_\mathrm{e}$ followed by traditional PL readout of the electron spin.
 The first readout cycle collapses the nuclear spin into an eigenstate, and ideally, each subsequent cycle polarizes the electron spin, but does not affect the nucleus, such that the photon counts from each readout window can be added.
 In reality, the number of cycles is limited by backaction from the measurement that eventually flips the nuclear spin.

 The initial demonstration by Jiang et al. \cite{Jiang2009} used a $^{13}$C nucleus with relatively strong coupling ($A_\parallel=\SI{14}{\mega\hertz}$).
 The map-and-measure procedure was repeated 30 times, improving the SNR by a factor of 2.2 compared to the traditional PL method.
 Subsequent improvements to the protocol, utilizing a $^{15}$N nuclear spin \cite{Lovchinsky2016}, resulted in an overall SNR boost by a factor of 6.8 after 500 cycles (Figure~\ref{fig:nuclear_assisted}c).
 This readout performance, used together with a sequence of quantum operations on the electron spin designed to sense weak oscillating magnetic fields from nuclear ensembles outside the diamond (Figure~\ref{fig:nuclear_assisted}b), enabled the detection of deuterated proteins on a diamond surface \cite{Lovchinsky2016}.

 The nuclear-assisted technique is technically demanding, requiring the application of complex quantum-control pulse sequences at both microwave and radio frequencies, precise alignment of an external dc magnetic field, and the identification or creation of an NV center with a suitably-coupled $^{13}$C or $^{15}$N (the natural isotopic abundance of $^{15}$N is 0.4\%).
 Furthermore, the time required for the C$_\mathrm{n}$NOT$_\mathrm{e}$ gate scales as $A_\parallel^{-1}$.
 This gate time introduces substantial overhead in the measurement, especially for weakly-coupled nuclei, limiting the measurement bandwidth and suppressing the sensitivity.
 On the other hand, more strongly-coupled nuclei suffer from unwanted spin-flips due to the nonsecular terms in Equation~(\ref{eq:Hhf}), limiting the number of cycles that can be performed and the achievable SNR. 

 For example, the ground-state hyperfine coupling to $^{14}$N is only $A_\parallel=\SI{2.16}{\mega\hertz}$, and the secular approximation holds (Figure \ref{fig:nuclear_assisted}a), whereas in the excited state, $A_\parallel \approx A_{\perp} \approx \SI{40}{\mega\hertz}$.
 Cycling through the excited state is unavoidable during the readout protocol, however, and the $A_\perp$ coupling severely limits the nuclear spin lifetime.
 At room-temperature, the flip-flop probability is maximized at the excited-state level anti-crossing (Lac) near \SI{500}{\Gauss} \cite{Neumann2009}.
 Interestingly, flip-flop transitions near the Lac can actually serve to increase the SNR, since a cascaded set of transitions allow for the spin-dependent PL contrast to persist for longer times, leading to a $\sqrt{3}$ increase in SNR \cite{Steiner2010}.
 Such cascaded transitions should produce sub-Poissonian noise \cite{DAnjou2017}, in which case the achievable SNR improvement might actually be somewhat larger.
 However, this technique only works within $\pm\SI{50}{\Gauss}$ of the excited sate Lac, and it requires both electron and nuclear control pulses.

\begin{figure*}[t]
\includegraphics[scale=1]{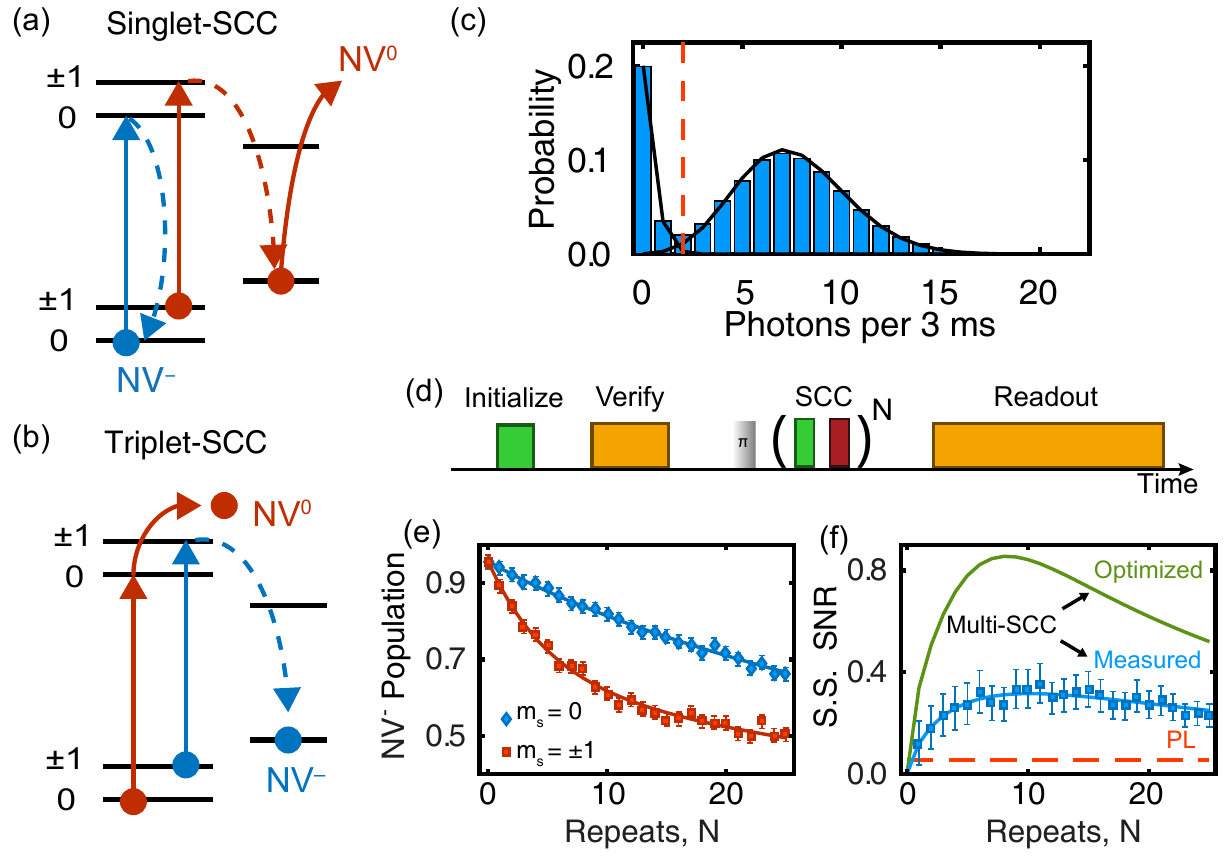}
\caption{\label{fig:scc} {Spin-to-charge conversion}. (\textbf{a},\textbf{b}) Schematics of the spin-dependent ionization pathways for singlet spin-to-charge conversion (S-SCC) and triplet-SCC (T-SCC), respectively. Solid lines represent laser induced transitions, while dashed lines represent decay transitions. (\textbf{c}) Histogram of photon counts during a \SI{3}{\milli\second} charge readout measurement with \SI{592}{\nano\meter} illumination \cite{Hopper2016}. (\textbf{d}) Timing diagram for the S-SCC protocol. (\textbf{e}) NV$^-$ population for different initial spin states as a function of the number of S-SCC repeats, \mbox{$N$ \cite{Hopper2016}}. (\textbf{f}) Single-shot (S.S.) SNR for S-SCC as a function of $N$ for the protocol as-demonstrated and for the optimal case assuming $100\%$ singlet ionization probability. The corresponding traditional-PL SNR is the dashed line at SNR = 0.055 \cite{Hopper2016}. Panels (\textbf{c}--\textbf{f}) are from \cite{Hopper2016}. Copyright 2016 by the American Physical Society.}
\end{figure*}

 Alternatively, at very high magnetic fields ($B>\SI{2500}{\Gauss}$), the large energy separation of spin eigenstates suppresses flip-flop interactions with $^{14}$N, as long as the field is precisely aligned to the NV-center symmetry axis.
 By operating at these fields, Neumann et al. \cite{Neumann2010} reached the single-shot readout regime for the $^{14}$N nuclear spin, with a fidelity of $92\%$.
 Subsequent analysis of the single-shot technique in the context of quantum sensing shows how the time-averaged SNR can be improved by an order of magnitude compared to traditional PL readout \cite{Haberle2017}.

 Despite their technical difficulty, nuclear-assisted readout protocols have been widely used in state-of-the-art demonstrations of single-NV quantum sensors \cite{Waldherr2012,Zaiser2016,Pfender2017,Aslam2017,Lovchinsky2016}.
Ideally, nuclear-assisted readout demands the following criteria: fast C$_\mathrm{n}$NOT$_\mathrm{e}$ operations to minimize measurement overhead, minimization of nonsecular components of the hyperfine Hamiltonian, and a nuclear spin with a long lifetime.
 These criteria are somewhat contradictory, in that fast gate operations require relatively strong coupling, which usually leads to larger nonsecular terms and shorter nuclear lifetimes.
 Nonetheless, they can be met in practice using any of the common nuclear species: $^{14}$N, $^{15}$N, or $^{13}$C.
 Application-specific experimental requirements often dominate the final selection.
 The primary physical limitation in most demonstrations remains the small, but non-zero, electron-nuclear flip-flop probability, especially in the NV center's excited state.
 These nonsecular terms can be reduced by selecting coupled $^{13}$C, which are closely aligned with the NV-center symmetry axis \cite{Dreau2013}.
 For the case of nuclear quantum \mbox{memories \cite{Maurer2012}}, uncontrolled transitions between the negatively- and neutrally-charged states of the NV center present further complications.
 Continued research into controlling these effects can help to extend the coupled nuclear spin's lifetime \cite{Maurer2012,Pfender2017a}.
 For quantum sensing applications, full consideration of the measurement-duration overhead (Equation~(\ref{eqn:time_avg_snr})) can help to optimize the sensitivity, especially for measurements on faster timescales.

\section{Spin-to-Charge Conversion \label{sec:scc}}
Whereas incomplete PL contrast and spin repolarization limit the fidelity of traditional spin measurements, the NV center's charge state can be measured optically with high precision, even at room temperature.
Given a means for mapping spin projections onto charge populations, or {spin to charge conversion} (SCC), charge measurements provide an alternate means to boost the spin-readout fidelity.
 SCC mechanisms are widely used in other spin-qubit platforms including quantum dots \cite{Elzerman2004} and silicon donors \cite{Morello2010}.
 In this section, we review two related mechanisms for all-optical SCC that exploit the NV center's ISC dynamics, and discuss how the tunability of the subsequent charge-readout process can be an advantage.

 High-SNR readout of the NV center's charge state was first demonstrated for single NV centers by Waldherr et al. \cite{Waldherr2011}, and the idea has since been extended to NV ensembles \cite{Jayakumar2016,Dhomkar2016} and nanodiamonds \cite{Hopper2018}.
 The charge readout mechanism depends on the energy difference between the zero phonon line (ZPL)
 optical transitions for the neutral (NV$^{0}$, ZPL at \SI{575}{\nano\meter}) and negative (NV$^{-}$, ZPL at \SI{637}{\nano\meter}) charge configurations, both of which are stable at room temperature.
 By selecting an excitation wavelength between these ZPLs, such as \SI{592}{\nano\meter}, only the NV$^-$ configuration is excited.
 When the optical power is tuned well below saturation, it is possible to detect more than one photon from NV$^-$ before an optically-induced charge transition to the dark NV$^0$ state occurs \cite{Waldherr2011a,Aslam2013}.

 The charge-readout SNR can be varied by changing the excitation power and readout duration.
 By using low powers and readout duration $>$$\SI{1}{\milli\second}$, single-shot charge fidelities exceeding 99$\%$ have been demonstrated for single NVs within photonic structures \cite{Shields2015,Hopper2016}.
 Figure~\ref{fig:scc}c shows an example of the photon-count histogram that results from a 3-ms-duration charge-readout measurement using \SI{592}{\nano\meter} light following initialization with a \SI{532}{\nano\meter} pulse.
 The clear separation of the count distribution into two Poissonian peaks is characteristic of high-fidelity readout, in this case with $\mathcal{F}=99.1\pm0.4$\% using the threshold of two photons shown by a dashed vertical line.

SCC can be achieved in two related ways, as shown in Figure~\ref{fig:scc}a,b.
Both techniques leverage the strong spin selectivity of the ISC from the NV center's $^3$E triplet excited state to the singlet manifold.
Following a single excitation event, a spin initially in $m_s=\pm1$ crosses to the singlet state with $\approx$50\% probability, whereas the $m_s=0$ state undergoes ISC only 5\% of the time \cite{Goldman2015a}.
Therefore, both techniques begin with a shelving step, consisting of a short, $<$20 ns, visible pulse of light that excites the triplet manifold with high probability.
After waiting for a time longer than the $^3$E excited-state lifetime (typically $\approx$\SI{20}{\nano\second}), a large fraction of the initial $m_s=\pm1$ spin population is stored in the metastable singlet ground state.
Next, an intense ionization pulse resonant with either the singlet absorption band (900--\SI{1042}{\nano\meter}, Figure~\ref{fig:scc}a) or the triplet absorption band (500--\SI{637}{\nano\meter}, Figure~\ref{fig:scc}b) is applied to ionize the singlet or triplet populations, respectively.
Hereafter, the methods will be referred to as singlet-SCC and triplet-SCC, depending on which manifold is ionized.

The two methods each have advantages and disadvantages.
Triplet-SCC relies on a highly efficient two-photon ionization process for the triplet using $\approx$600--\SI{637}{\nano\meter} light \cite{Aslam2013,Shields2015}.
This can be the same color used for both the shelving step and subsequent charge readout \cite{Hopper2016}, which simplifies experiments.
However, the triplet-SCC efficiency is ultimately limited by the 50\% ISC probability for $m_s=\pm1$ spin states, since any population that remains in the triplet after the shelving step is ionized.
\mbox{Shields et al. \cite{Shields2015}} essentially reached the practical limit for this technique, demonstrating a single-shot $\mathcal{F}=67\%$, corresponding to an SNR increased by a factor of 4.1 over traditional PL (the SNR ratio in this case is limited by the high collection efficiency in this experiment).

Singlet-SCC, on the other hand, leaves the triplet population unaffected, and the shelve-ionize procedure can be rapidly repeated as shown in Figure~\ref{fig:scc}d, ideally to reach the maximum SCC efficiency given by the $\approx$10:1 spin-dependent ISC branching ratio.
Figure~\ref{fig:scc}e,f shows how the spin-dependent charge contrast and corresponding single-shot SNR vary with the number of repeats, $N$.
Drawbacks of this approach include the need for both visible and near-infrared optical beams, and the small optical cross-section for the singlet optical transition \cite{Acosta2010}, which necessitates a high-intensity near-infrared pulse to achieve 100\% ionization efficiency.
For the data shown in Figure~\ref{fig:scc}e,f, the singlet ionization probability was only 25\%, and the singlet-SCC protocol achieved a maximum $\mathcal{F}=62\%$, corresponding to an SNR increase by a factor of 5.8 over traditional PL \cite{Hopper2016}.
The infrared pulses used by \mbox{Hopper et al. \cite{Hopper2016}} were derived from a supercontinuum laser, bandpass filtered to 900--\SI{1000}{\nano\meter}, with an average picosecond pulse energy of \SI{2}{\nano\joule}. 
Since the ionization rate depends quadratically on pulse energy, increasing the pulse energy by an order of magnitude should lead to ionization probabilities exceeding 99\%.
Assuming 100\% ionization can be achieved using higher optical pulse energies, Figure~\ref{fig:scc}f shows how the singlet-SCC protocol with $N=8$ repeats can achieve SNR $>$ 0.84, corresponding to $\mathcal{F}>75$\% and an increase over traditional PL by a factor of 15.

Recently, the benefits from SCC have been explored in materials platforms more suited to sensing such as NVs beneath planar surfaces \cite{Jaskula}, shallow NVs in nanopillars \cite{Ariyaratne2018}, and NV ensembles within type-Ib nanodiamonds \cite{Hopper2018}.
 These promising results suggest that SCC can boost the performance of myriad applications.


\section{Photocurrent Readout
\label{sec:photocurrent}}
 The free electrons and holes produced from photoionization can be utilized as an observable of the NV center's spin state.
 By taking advantage of the same spin-dependent ionization phenomenon that enables SCC (see Section~\ref{sec:scc}), spin-dependent photocurrents can be produced.
 Although it is still in the early stages, electrical readout potentially offers improvements in speed, together with a scalable means for integrating many NV devices on a chip with high density.
 In this section, we overview the recent advances in photocurrent spin-readout and discuss future directions of research.
 
 Photocurrent readout is possible due to the propensity for the $m_s=\pm1$ spin states to be shelved into the singlet manifold \cite{Bourgeois2015}, protecting them from ionization for roughly the singlet lifetime ($\approx$$\SI{200}{\nano\second}$). 
 Meanwhile, if optical intensities well above the saturation value drive the triplet optical transition, rapid ionization and recombination processes generate free electrons and holes, respectively, for the initial $m_s=0$ spin projection.
 The goal of photocurrent readout is two-fold: to maximize the number of free carriers produced within the \SI{200}{\nano\second} shelving time through the use of very high intensity \SI{532}{\nano\meter} illumination, and to efficiently collect and amplify the current while avoiding unwanted noise. 
 Initial experiments demonstrated electrical detection of continuous-wave electron spin resonance for ensembles of NV centers \cite{Brenneis2015,Bourgeois2015}.
 Recent advances in device design, lock-in detection, pulsed measurements, and multi-color pump beams have lead to improved contrast \cite{Bourgeois2017,Hrubesch2017} and a current detection limit of only five NV centers \cite{Gulka2017}.

 \begin{figure*}[t]
 \includegraphics[scale=1]{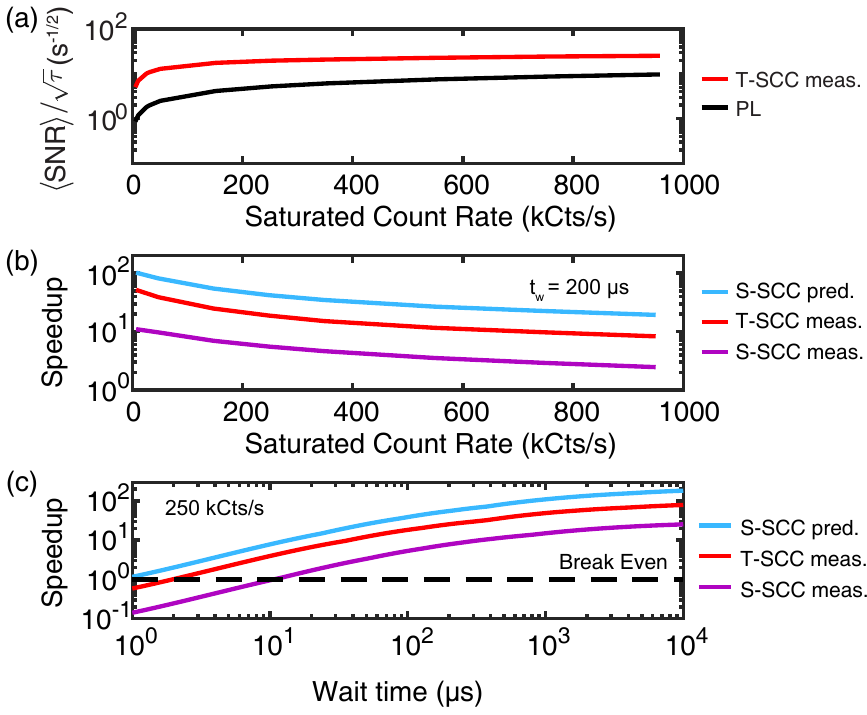}
 \caption{\label{fig:speedup} {Quantifying SCC improvements in experiments}. (\textbf{a}) Time averaged SNR scaled by $\sqrt{\tau}$, for the traditional PL and triplet-SCC protocols as a function of saturation green-illumination count rate, assuming $t_W= \SI{200}{\micro\second}$. The T-SCC SNR is numerically calculated using the model in \citep{Shields2015}. (\textbf{b}) Speedup comparison for the various SCC techniques as a function of green saturation count rate, assuming $t_W=\SI{200}{\micro\second}$. (\textbf{c}) Speedup comparison as a function of $t_W$, assuming a green saturation count rate of \SI{250}{\kilo\counts\per\second}. The dashed line indicates the ``break-even'' point, where SCC provides a more efficient readout than traditional PL. The speedup in ({b},{c}) is calculated using data reported in \cite{Shields2015, Hopper2016}.}
\end{figure*}
 
 Looking forward, the detection of spin-dependent photocurrent from a single NV center remains an outstanding challenge. 
 Such experiments will require careful analysis of the entire electronic noise budget and materials optimization to remove background photocurrents due to substitutional nitrogen and other defects \cite{Heremans2009}.
 Due to the similarity with optical SCC, certain aspects of the SCC pulse sequences could be adapted to electrical readout.
 Systematic investigations of the optimal shelf and ionization colors, durations, and powers could further increase the SNR from photocurrent based readout.
 Despite these challenges, electrical detection of NV center spin states has enormous potential for developing integrated sensors and devices.

\section{Accounting for Measurement Overhead
\label{sec:measurement_overhead}}
 When averaging is required, the time spent initializing and measuring reduces the achievable time-averaged SNR and sensitivity (see Equations~(\ref{eqn:time_avg_snr}) and (\ref{eqn:sensitivity}), respectively).
 Since traditional PL readout consists of a short duration of a few hundred nanoseconds, the measurement overhead is usually a fixed penalty with little room for improvement.
 However, more advanced readout protocols such as low-temperature, nuclear-assisted, and SCC techniques feature measurement times that can be comparable to or longer than typical spin evolution times.
 In this case, the measurement overhead becomes a major factor, but there is also added flexibility in designing protocols since the single-shot SNR typically depends on the measurement duration.
 Optimizing the trade-off between the number of experimental repeats and single-shot SNR can result in drastic improvements in time-averaged SNR.
 Here, we describe the process for optimizing the measurement overhead in the context of SCC readout, using a model that can be directly adapted to nuclear-assisted readout \cite{Haberle2017} and \mbox{low-temperature readout}.


An arbitrary NV-center measurement can be broken up into three times: the initialization time, $t_I$, the wait time, $t_W$, and the readout time $t_R$, such that the total duration of a single measurement is:

\begin{equation}
\tau = t_I + t_W + t_R.
\end{equation}
Following from Equation~(\ref{eqn:time_avg_snr}), the time-averaged SNR is given by:

\begin{equation}
\langle\mathrm{SNR}\rangle_{\mathrm{SCC}} = \sqrt{\frac{T}{\tau}}\mathrm{SNR}\left(t_R, P_R\right),
\end{equation}
where $T$ is the total integration time, and $\mathrm{SNR}(t_R, P_R)$ is the single-shot SNR as a function of $t_R$ and the optical power, $P_R$.
The single-shot SNR can be experimentally calibrated for various settings of $(t_R,P_R)$, or it can be calculated using a numerical model of the charge readout process accounting for ionization and recombination processes that become important as $P_R$ increases \cite{Shields2015}.
Given desired experimental settings for $t_I$ and $t_W$, optimal readout parameters can be chosen to maximize $\langle \mathrm{SNR}\rangle_\mathrm{SCC}$.
In some cases, it can also be beneficial to optimize over $t_W$ and $t_I$, e.g., for sensing applications by using a suitable formulation for the field sensitivity (Equation~(\ref{eqn:sensitivity})) that accounts for the signal amplitude as a function of $t_W$, as well as the time-averaged SNR.
In general, experiments with longer wait times such as dynamical decoupling sequences for ac field sensing, $T_1$ measurements, and controlled interactions with nuclear spins stand to gain the largest performance improvements from SCC.

 A useful metric to quantify the SCC performance is the speedup,

\begin{equation}
 \textrm{Speedup} = \frac{T_{\textrm{PL}}}{T_\textrm{SCC}} = \frac{\tau_{\textrm{PL}}}{\tau_{\textrm{SCC}}}\left(\frac{\textrm{SNR}_{\textrm{SCC}}}{\textrm{SNR}_{\textrm{PL}}}\right)^2,
\end{equation}
 where $\tau_\textrm{SCC}$, $T_\textrm{SCC}$, SNR$_{\textrm{SCC}}$ ($\tau_\textrm{PL}$, $T_\textrm{PL}$, and SNR$_{\textrm{PL}}$) represent the measurement-cycle duration, total integration time, and single-shot SNR for SCC (PL), respectively. 
 The speedup quantifies the reduction in total integration time required to achieve a desired time-averaged SNR when using SCC as opposed to traditional PL.
 A speedup $>$1 implies that SCC is more efficient than traditional PL.
 When the time-averaged SNR is optimized over $t_R$ and $P_R$ as a function of $t_W$, the value of $t_W$ at which the speedup exceeds unity is referred to as the break-even time.

 All of these quantities need to be calculated or measured for a given experimental setting accounting for the photon collection efficiency, SCC efficiency, etc.
 Figure~\ref{fig:speedup} gives an example of such optimization calculations, showing how the time-averaged SNR for PL and SCC protocols depend on photon count rate, and the corresponding speedup as a function of the count rate and $t_W$.
 The flexibility of optimizing the SCC readout settings can offer impressive gains; both singlet-SCC and triplet-SCC exhibit order-of-magnitude speedups for experimentally relevant wait times, and the optimized singlet-SCC protocol approaches a 100-fold speedup.

 The application of the measurement overhead optimization framework could prove beneficial in drastically speeding up nuclear-assisted and low-temperature readout experiments.
 Future extensions of this technique could focus on additionally optimizing initialization times, where in-the-loop feedback is used to verify proper charge, nuclear, or electron states.

\section{Real-Time Signal Processing Techniques \label{sec:signal_processing}}

 The growing number of spin-readout techniques discussed in the previous sections all aim to overcome photon shot noise by increasing the number of photons that can be recorded in each measurement cycle.
 In this situation, it is beneficial to leverage signal-processing techniques that make use of the time-of-arrival information of each photon, as opposed to simply summing the total number of detections in a fixed time window.
 This approach can even be applied to traditional PL spin readout, with an SNR improvement of 7$\%$ \cite{Gupta2016}.
 Much larger gains can be achieved when each measurement yields multiple photons.
 Together with low-temperature resonance-fluorescence readout, real-time detection protocols have been essential for the achievement of heralded entanglement \cite{Bernien2013} and partial measurements \cite{Blok2014}, since they boost the readout fidelity while reducing unwanted backaction.
 Similarly, hidden Markov models can improve the performance of room-temperature, single-shot charge-state readout \cite{DAnjou2016}.
 These results imply that significant improvements should be achievable for room-temperature applications using real-time signal processing in conjunction with nuclear-assisted or SCC readout protocols.
 With the increasing number of related demonstrations and the larger quantum-information community's emphasis on open-source tools \cite{artiq,Qudi2017}, the technological hurdles of implementing real-time analysis will be overcome.

\section{Discussion \label{sec:discussion}}
\begin{figure*}[t]
\includegraphics{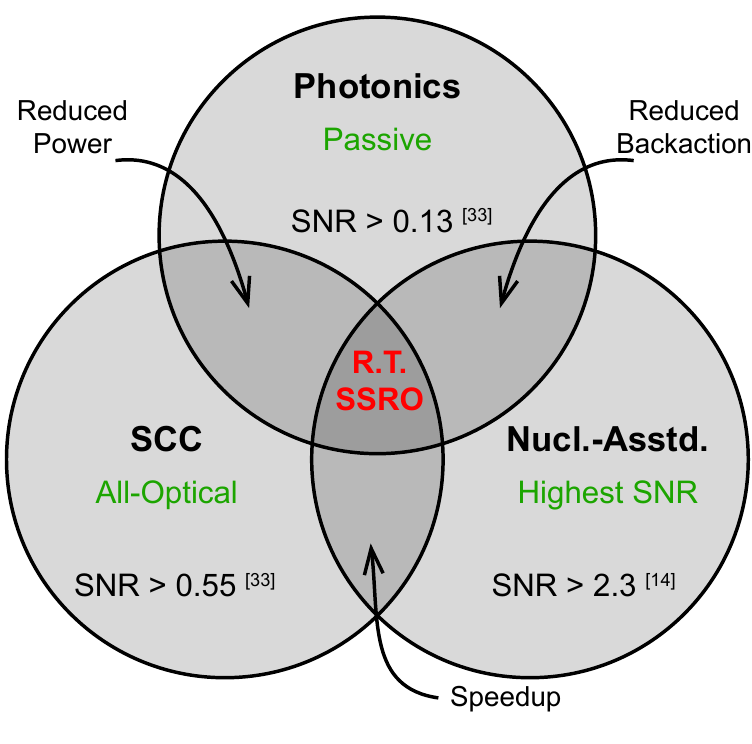}
\caption{\label{fig:summary}
Complementary approaches for enhanced spin readout. Existing techniques have advantages for particular applications. Future research can consider the potential for combining multiple techniques in order to achieve fast, high-fidelity, single-shot readout (SSRO) of the NV center's electron spin at room temperature. The highest reported traditional PL SNR, as well as the SCC SNR, are from Shields et al. \cite{Shields2015}. The highest nuclear assisted SNR is from Neumann et al. \cite{Neumann2010}.}
\end{figure*}

Although the techniques and approaches discussed in this review have mostly developed independently, they are not mutually exclusive.
 Figure~\ref{fig:summary} depicts the key advantages of room-temperature approaches based on photonics engineering, SCC, and nuclear quantum logic, including the current state-of-the-art SNR achieved in each case.
 In many cases, combinations of multiple techniques could overcome existing limitations and provide significant improvements in spin-readout SNR.

 In addition to providing a Purcell enhancement of the NV center's PL emission rate, as discussed in Section \ref{sec:radiative_lifetime}, plasmonic antennae can also be designed to enhance optical absorption from incident radiation fields.
 Enhancing absorption is especially important for biological applications, where background autofluorescence and phototoxicity associated with incident \SI{532}{\nano\meter} light limit the optical intensity that can be applied and the achievable SNR.
 Absorption-enhancing plasmonic structures similar to those currently used to improve thin-film solar cells \cite{Lim2007} could reduce the input power, and even enable up-converting schemes for biological applications based on two-photon absorption at near-infrared wavelengths \cite{Tse-LuenWee2007}.
 Similarly, the singlet-SCC protocol would also benefit from absorption enhancements at near-infrared wavelengths, since its fidelity is currently limited by incomplete ionization of the singlet manifold due to a small singlet absorption cross-section \cite{Hopper2016}.
 Photocurrent-based readout techniques would also stand to gain from absorption enhancement, due to the high power requirements for rapidly changing of the charge state.

 Real-time signal-processing techniques present immediate benefits to applications requiring non-destructive, single-shot readout, and they can also improve the performance of sensing applications.
 For example, real-time analysis reduces the average measurement time for charge detection by a factor of two \cite{DAnjou2016}, and the exact same hardware can verify proper charge-state initialization.
 Purifying measurements on the basis of the NV$^-$ charge state leads to reduced \mbox{noise \cite{Waldherr2011,Hopper2016}} and is compatible with sensing schemes \cite{Shields2015}.
 Usually, such purification is performed using post-selection, but a combination of real-time charge verification and dynamic readout should yield more than a 50\% improvement in SNR for both SCC and nuclear-assisted \mbox{readout protocols}.

 The key limitation on readout fidelity for nuclear-assisted readout protocols is the effective nuclear spin lifetime, which is reduced by cycling between the electron spin's ground and excited states \cite{Neumann2010}.
 The current solution to this problem is to work at very high magnetic fields where hyperfine-induced bit-flip errors are reduced, but radiative lifetime engineering that reduces the excited-state lifetime presents an alternative approach.
 Achievable radiative rate enhancements of up to two orders of magnitude \cite{Bogdanov2017,Karamlou2018,Bogdanov2017a} could increase the nuclear-spin $T_1$ by a similar factor, enabling higher-fidelity measurements or relaxing the field constraints.
 Alternatively, combinations of nuclear-assisted and SCC readout protocols could take advantage of SCC's wide measurement tunability to limit the total number of optical cycles while maintaining overall SNR.
 Either of these approaches could further reduce the readout errors for room-temperature, single-shot readout to the level of a few percent.

\section{Conclusions \label{sec:conclusion}}
 In reviewing the state-of-the-art for optical spin readout of the diamond NV center, we hope to spur further advances in this field and encourage the adoption of more sophisticated techniques in future experiments.
 In general, enhancements result from increasing the number of detected photons, either by directly modifying the photon emission rate or by mapping the electron spin state onto a longer-lived observable.
 Each of these approaches has advantages for particular applications, with varying technical requirements in terms of fabrication technology and experimental complexity.
 \mbox{For this} reason, there is no clear front runner, and it is likely that all of these techniques will continue to improve in future experiments.
 To date, there has been little exploration into how different techniques can be combined with each other (Figure~\ref{fig:summary}) or enhanced using real-time signal processing.
 Since spin-readout performance impacts nearly every application of NV centers for quantum science and technology, these questions are compelling avenues for future research.

\vspace{6pt}


\acknowledgments{The authors thank Richard Grote for preparing the solid immersion lens graphic. This work was supported by the National Science Foundation through a CAREER Award (EECS-1553511) and the University of Pennsylvania Materials Research Science and Engineering Center (MRSEC) (DMR-1720530).}

\appendix
\section{Spin-Readout Noise Calculations \label{appendix:sigmaR_calculations}}
 Following Equation~(\ref{eqn:angle_deviation}), in order to calculate the spin-readout noise, $\sigma_R$, we must identify the signal's dependence on angle, $\frac{\partial \langle S\rangle}{\partial\theta}$, and the standard deviation, $\sigma_S$.
 The standard deviation differs from the common form for Poisson or binomial random variables due to the signal consisting of the {sum} of two random variables with different weights, and therefore, we use the general expression for \mbox{the variance},
\begin{equation}
\sigma_S^2 = \langle S^2 \rangle - \langle S \rangle^2.
\label{eqn:variance}
\end{equation}
In this Appendix, we derive expressions for $\sigma_R$ corresponding to photon summation and thresholding signals.

\subsection{Photon Summation}

 For photon summation, the signal (Equation~(\ref{eqn:generic_signal})) directly reflects the number of detected photons:
\begin{equation}
\langle S \rangle = \cos^2\left(\frac{\theta}{2}\right)\alpha_0 + \sin^2\left(\frac{\theta}{2}\right)\alpha_1.
\label{eqn:poisson_signal}
\end{equation}
 From this expression, we can directly derive the signal variation,
\begin{equation}
\frac{\partial \langle S \rangle}{\partial \theta} = \frac{1}{2}\sin(\theta)(\alpha_0-\alpha_1),
\label{eqn:poisson_signal_deviation}
\end{equation}
and the expectation value of the signal squared,
\begin{equation}
\langle S^2 \rangle = \cos^2\left(\frac{\theta}{2}\right)(\alpha_0^2 + \alpha_0) + \sin^2\left(\frac{\theta}{2}\right)(\alpha_1^2+\alpha_1).
\label{eqn:poisson_signal_squared}
\end{equation}
 The latter expression uses the fact that the expected value of $X^2$ from a Poisson distribution $P(X;\lambda)$ with mean-value $\lambda$ is $\langle X^2\rangle=\lambda^2+\lambda$.
 By combining Equations~(\ref{eqn:variance})--(\ref{eqn:poisson_signal_squared}) with Equation~(\ref{eqn:projection_noise}),
we arrive at the following general expression for the spin-readout noise:
\begin{equation}
\sigma_R^{\textrm{Photon}} = \frac{\sqrt{\frac{1}{4}\sin^2(\theta)(\alpha_0-\alpha_1)^2 + \cos^2\left(\frac{\theta}{2}\right)\alpha_0 + \sin^2\left(\frac{\theta}{2}\right)\alpha_1}}{\frac{1}{2}\sin(\theta)(\alpha_0-\alpha_1)}.
\label{eqn:full_photon_readout_noise}
\end{equation}
 Assuming an equally-weighted superposition state $(\theta=\pi/2)$, Equation~(\ref{eqn:full_photon_readout_noise}) reduces to the form reported in Equation~(\ref{eqn:photon_readout_noise}) of the main text.

\subsection{Thresholding}

 In the case of thresholding, the signal results from the binary values of the measurement, where we assume that $S=1$ corresponds to the identification of the zero spin state (Equation~(\ref{eqn:threshold_mean})), \mbox{and therefore:}
\begin{equation}
\langle S \rangle = \cos^2\left(\frac{\theta}{2}\right)p_{0|0} + \sin^2\left(\frac{\theta}{2}\right)p_{0|1}.
\label{eqn:threshold_signal}
\end{equation}
 As before, we use this expression to calculate the signal variation,
\begin{equation}
\frac{\partial \langle S \rangle}{\partial \theta} = \frac{1}{2}\sin(\theta)(p_{0|0} - p_{0|1}),
\label{eqn:threshold_signal_deviation}
\end{equation}
and the mean of the signal squared,
\begin{equation}
\langle S^2 \rangle = \cos^2\left(\frac{\theta}{2}\right)p_{0|0} + \sin^2\left(\frac{\theta}{2}\right)p_{0|1}.
\label{eqn:threshold_signal_squared}
\end{equation}
 
 These expressions yield the following general form for the spin-readout noise associated \mbox{with thresholding},
\begin{widetext}
\begin{equation}
\sigma_R^{\textrm{Threshold}} = \frac{\sqrt{\left(\cos^2\left(\frac{\theta}{2}\right)p_{0|0} - \sin^2\left(\frac{\theta}{2}\right)p_{0|1}\right)^2
+ p_{0|0}\left( \cos^2\left(\frac{\theta}{2}\right) - 2\cos^4\left(\frac{\theta}{2}\right)\right)
+ p_{0|1}\left(\sin^2\left(\frac{\theta}{2}\right) - \sin^4\left(\frac{\theta}{2}\right)\right)
}}{\frac{1}{2}\sin(\theta)(p_{0|0}-p_{0|1})},
\label{eqn:full_threshold_readout_noise}
\end{equation}
\end{widetext}
 Assuming an equally-weighted superposition state $(\theta=\pi/2)$, Equation~(\ref{eqn:full_threshold_readout_noise}) reduces to the form reported in Equation~(\ref{eqn:threshold_readout_noise}) of the main text.


%

\begin{thebibliography}{}
\providecommand{\natexlab}[1]{#1}

\bibitem[Gambetta \em{et~al.}(2017)Gambetta, Chow, and Steffen]{Gambetta2017}
Gambetta, J.M.; Chow, J.M.; Steffen, M.
\newblock {Building logical qubits in a superconducting quantum computing
 system}.
\newblock {\em npj Quantum Inf.} {\bf 2017}, {\em 3},~2.

\bibitem[Brown \em{et~al.}(2016)Brown, Kim, and Monroe]{Brown2016}
Brown, K.R.; Kim, J.; Monroe, C.
\newblock {Co-designing a scalable quantum computer with trapped atomic ions}.
\newblock {\em \mbox{npj Quantum Inf.}} {\bf 2016}, {\em 2},~16034.

\bibitem[Silverstone \em{et~al.}(2016)Silverstone, Bonneau, O'Brien, and
 Thompson]{Silverstone2016}
Silverstone, J.W.; Bonneau, D.; O'Brien, J.L.; Thompson, M.G.
\newblock {Silicon Quantum Photonics}.
\newblock {\em IEEE J. Sel. Top. Quantum Electron.} {\bf
 2016}, {\em 22},~390--402.

\bibitem[Awschalom \em{et~al.}(2013)Awschalom, Bassett, Dzurak, Hu, and
 Petta]{Awschalom2013}
Awschalom, D.D.; Bassett, L.C.; Dzurak, A.S.; Hu, E.L.; Petta, J.R.
\newblock {Quantum Spintronics: Engineering and Manipulating Atom-Like Spins in
 Semiconductors}.
\newblock {\em Science} {\bf 2013}, {\em 339},~1174--1179.

\bibitem[DiVincenzo(2000)]{DiVincenzo2000}
DiVincenzo, D.P.
\newblock {The Physical Implementation of Quantum Computation}.
\newblock {\em Fortschr. Phys.} {\bf 2000}, {\em 48},~771--783.

\bibitem[Nielsen and Chuang(2000)]{nielsen2000}
Nielsen, M.A.; Chuang, I.
\newblock {\em Quantum computation and quantum communication}. Cambridge University Press: Cambridge, England \textbf{2010}.

\bibitem[Clerk \em{et~al.}(2010)Clerk, Devoret, Girvin, Marquardt, and
 Schoelkopf]{Clerk2010}
Clerk, A.A.; Devoret, M.H.; Girvin, S.M.; Marquardt, F.; Schoelkopf, R.J.
\newblock {Introduction to quantum noise, measurement, and amplification}.
\newblock {\em Rev. Mod. Phys.} {\bf 2010}, {\em 82},~1155--1208.

\bibitem[Heremans \em{et~al.}(2016)Heremans, Yale, and Awschalom]{Heremans2016}
Heremans, F.J.; Yale, C.G.; Awschalom, D.D.
\newblock {Control of Spin Defects in Wide-Bandgap Semiconductors for Quantum
 Technologies}.
\newblock {\em Proc. IEEE} {\bf 2016}, {\em 104},~2009--2023.

\bibitem[Doherty \em{et~al.}(2013)Doherty, Manson, Delaney, Jelezko, Wrachtrup,
 and Hollenberg]{Doherty2013}
Doherty, M.W.; Manson, N.B.; Delaney, P.; Jelezko, F.; Wrachtrup, J.;
 Hollenberg, L.C.L.
\newblock {The nitrogen-vacancy colour centre in diamond}.
\newblock {\em Phys. Rep.} {\bf 2013}, {\em 528},~1--45.

\bibitem[Dutt \em{et~al.}(2007)Dutt, Childress, Jiang, Togan, Maze, Jelezko,
 Zibrov, Hemmer, and Lukin]{Dutt2007}
Dutt, M.V.G.; Childress, L.; Jiang, L.; Togan, E.; Maze, J.; Jelezko, F.;
 Zibrov, A.S.; Hemmer, P.R.; Lukin,~M.D.
\newblock {Quantum Register Based on Individual Electronic and Nuclear Spin
 Qubits in Diamond}.
\newblock {\em Science} {\bf 2007}, {\em 316},~1312--1316.

\bibitem[Maurer \em{et~al.}(2012)Maurer, Kucsko, Latta, Jiang, Yao, Bennett,
 Pastawski, Hunger, Chisholm, Markham, Twitchen, Cirac, and Lukin]{Maurer2012}
Maurer, P.C.; Kucsko, G.; Latta, C.; Jiang, L.; Yao, N.Y.; Bennett, S.D.;
 Pastawski, F.; Hunger, D.; Chisholm,~N.; Markham, M.; et al.
\newblock {Room-Temperature Quantum Bit Memory Exceeding One Second}.
\newblock {\em Science} {\bf 2012}, {\em 336},~1283 LP--1286.

\bibitem[Pfender \em{et~al.}(2017)Pfender, Aslam, Simon, Antonov, Thiering,
 Burk, {F{\'{a}}varo de Oliveira}, Denisenko, Fedder, Meijer, Garrido, Gali,
 Teraji, Isoya, Doherty, Alkauskas, Gallo, Gr{\"{u}}neis, Neumann, and
 Wrachtrup]{Pfender2017a}
Pfender, M.; Aslam, N.; Simon, P.; Antonov, D.; Thiering, G.; Burk, S.;
 {F{\'{a}}varo de Oliveira}, F.; Denisenko, A.; Fedder, H.; Meijer, J.; et al.
\newblock {Protecting a Diamond Quantum Memory by Charge State Control}.
\newblock {\em Nano Lett.} {\bf 2017}, {\em 17},~5931--5937.

\bibitem[Childress \em{et~al.}(2006)Childress, {Gurudev Dutt}, Taylor, Zibrov,
 Jelezko, Wrachtrup, Hemmer, and Lukin]{Childress2006}
Childress, L.; {Gurudev Dutt}, M.V.; Taylor, J.M.; Zibrov, A.S.; Jelezko, F.;
 Wrachtrup, J.; Hemmer, P.R.; Lukin, M.D.
\newblock {Coherent Dynamics of Coupled Electron and Nuclear Spin Qubits in
 Diamond}.
\newblock {\em Science} {\bf 2006}, {\em 314},~281--285.

\bibitem[Neumann \em{et~al.}(2010)Neumann, Beck, Steiner, Rempp, Fedder,
 Hemmer, Wrachtrup, and Jelezko]{Neumann2010}
Neumann, P.; Beck, J.; Steiner, M.; Rempp, F.; Fedder, H.; Hemmer, P.R.;
 Wrachtrup, J.; Jelezko, F.
\newblock {Single-Shot Readout of a Single Nuclear Spin}.
\newblock {\em Science} {\bf 2010}, {\em 329},~542--544.

\bibitem[Liu \em{et~al.}(2017)Liu, Xing, Ma, Wang, Li, Po, Zhang, Fan, Liu, and
 Pan]{Liu2017}
Liu, G.Q.; Xing, J.; Ma, W.L.; Wang, P.; Li, C.H.; Po, H.C.; Zhang, Y.R.; Fan,
 H.; Liu, R.B.; Pan, X.Y.
\newblock {Single-Shot Readout of a Nuclear Spin Weakly Coupled to a
 Nitrogen-Vacancy Center at Room Temperature}.
\newblock {\em \mbox{Phys. Rev. Lett.}} {\bf 2017}, {\em 118},~150504.

\bibitem[Casola \em{et~al.}(2018)Casola, van~der Sar, and Yacoby]{Casola2018}
Casola, F.; van~der Sar, T.; Yacoby, A.
\newblock {Probing condensed matter physics with magnetometry based on
 nitrogen-vacancy centres in diamond}.
\newblock {\em Nat. Rev. Mater.} {\bf 2018}, {\em 3},~17088.

\bibitem[Lovchinsky \em{et~al.}(2016)Lovchinsky, Sushkov, Urbach, de~Leon,
 Choi, {De Greve}, Evans, Gertner, Bersin, M{\"{u}}ller, McGuinness, Jelezko,
 Walsworth, Park, and Lukin]{Lovchinsky2016}
Lovchinsky, I.; Sushkov, A.O.; Urbach, E.; de~Leon, N.P.; Choi, S.; {De Greve},
 K.; Evans, R.; Gertner, R.; Bersin, E.; M{\"{u}}ller, C.; et al.
\newblock {Nuclear magnetic resonance detection and spectroscopy of single
 proteins using quantum logic}.
\newblock {\em Science} {\bf 2016}, {\em 351},~836--841.

\bibitem[Aslam \em{et~al.}(2017)Aslam, Pfender, Neumann, Reuter, Zappe,
 F{\'a}varo~de Oliveira, Denisenko, Sumiya, Onoda, Isoya, and
 Wrachtrup]{Aslam2017}
Aslam, N.; Pfender, M.; Neumann, P.; Reuter, R.; Zappe, A.; F{\'a}varo~de
 Oliveira, F.; Denisenko, A.; Sumiya,~H.; Onoda, S.; Isoya, J.; et al.
\newblock Nanoscale nuclear magnetic resonance with chemical resolution.
\newblock {\em Science} {\bf 2017}, \emph{357}, 67--71.

\bibitem[Arcizet \em{et~al.}(2011)Arcizet, Jacques, Siria, Poncharal, Vincent,
 and Seidelin]{Arcizet2011}
Arcizet, O.; Jacques, V.; Siria, A.; Poncharal, P.; Vincent, P.; Seidelin, S.
\newblock {A single nitrogen-vacancy defect coupled to a nanomechanical
 oscillator}.
\newblock {\em Nat. Phys.} {\bf 2011}, {\em 7},~879--883.

\bibitem[Hensen \em{et~al.}(2015)Hensen, Bernien, Dr{\'{e}}au, Reiserer, Kalb,
 Blok, Ruitenberg, Vermeulen, Schouten, Abell{\'{a}}n, Amaya, Pruneri,
 Mitchell, Markham, Twitchen, Elkouss, Wehner, Taminiau, and
 Hanson]{Hensen2015}
Hensen, B.; Bernien, H.; Dr{\'{e}}au, A.E.; Reiserer, A.; Kalb, N.; Blok, M.S.;
 Ruitenberg, J.; Vermeulen, R.F.L.; Schouten, R.N.; Abell{\'{a}}n, C.; et al.
\newblock {Loophole-free Bell inequality violation using electron spins
 separated by 1.3 kilometres}.
\newblock {\em Nature} {\bf 2015}, {\em 526},~682.

\bibitem[Degen \em{et~al.}(2017)Degen, Reinhard, and Cappellaro]{Degen2017}
Degen, C.L.; Reinhard, F.; Cappellaro, P.
\newblock {Quantum sensing}.
\newblock {\em Rev. Mod. Phys.} {\bf 2017}, {\em 89},~35002.

\bibitem[Rondin \em{et~al.}(2014)Rondin, Tetienne, Hingant, Roch, Maletinsky,
 and Jacques]{Rondin2014}
Rondin, L.; Tetienne, J.P.; Hingant, T.; Roch, J.F.; Maletinsky, P.; Jacques,
 V.
\newblock {Magnetometry with nitrogen-vacancy defects in diamond}.
\newblock {\em Rep. Prog. Phys.} {\bf 2014}, {\em 77},~56503.

\bibitem[Schirhagl \em{et~al.}(2014)Schirhagl, Chang, Loretz, and
 Degen]{Schirhagl2014}
Schirhagl, R.; Chang, K.; Loretz, M.; Degen, C.L.
\newblock {Nitrogen-vacancy centers in diamond: nanoscale sensors for physics
 and biology}.
\newblock {\em Annu. Rev. Phys. Chem.} {\bf 2014}, {\em 65},~83--105.

\bibitem[Schr{\"{o}}der \em{et~al.}(2016)Schr{\"{o}}der, Mouradian, Zheng,
 Trusheim, Walsh, Chen, Li, Bayn, and Englund]{Schroder:16}
Schr{\"{o}}der, T.; Mouradian, S.L.; Zheng, J.; Trusheim, M.E.; Walsh, M.;
 Chen, E.H.; Li, L.; Bayn, I.; Englund, D.
\newblock {Quantum nanophotonics in diamond}.
\newblock {\em J. Opt. Soc. Am. B} {\bf 2016}, {\em 33},~B65--B83.

\bibitem[McDonough and Whalen(1995)]{McDonough1995}
McDonough, R.N.; Whalen, A.D.
\newblock {\em {Detection of Signals in Noise}}; Academic Press: Cambridge, MA, USA, 1995.

\bibitem[Vijay \em{et~al.}(2011)Vijay, Slichter, and Siddiqi]{Vijay2011}
Vijay, R.; Slichter, D.H.; Siddiqi, I.
\newblock {Observation of Quantum Jumps in a Superconducting Artificial Atom}.
\newblock {\em Phys. Rev. Lett.} {\bf 2011}, {\em 106},~110502.

\bibitem[Taylor \em{et~al.}(2008)Taylor, Cappellaro, Childress, Jiang, Budker,
 Hemmer, Yacoby, Walsworth, and Lukin]{Taylor2008}
Taylor, J.M.; Cappellaro, P.; Childress, L.; Jiang, L.; Budker, D.; Hemmer,
 P.R.; Yacoby, A.; Walsworth, R.; Lukin, M.D.
\newblock {High-sensitivity diamond magnetometer with nanoscale resolution}.
\newblock {\em Nat. Phys.} {\bf 2008}, {\em 4},~810--816.

\bibitem[D'Anjou and Coish(2014)]{DAnjou2014}
D'Anjou, B.; Coish, W.A.
\newblock {Optimal post-processing for a generic single-shot qubit readout}.
\newblock {\em Phys. Rev. A} {\bf 2014}, {\em 89},~12313.

\bibitem[D'Anjou \em{et~al.}(2016)D'Anjou, Kuret, Childress, and
 Coish]{DAnjou2016}
D'Anjou, B.; Kuret, L.; Childress, L.; Coish, W.A.
\newblock {Maximal Adaptive-Decision Speedups in Quantum-State Readout}.
\newblock {\em Phys. Rev. X} {\bf 2016}, {\em 6},~011017.

\bibitem[Robledo \em{et~al.}(2011)Robledo, Childress, Bernien, Hensen,
 Alkemade, and Hanson]{Robledo2011}
Robledo, L.; Childress, L.; Bernien, H.; Hensen, B.; Alkemade, P.F.A.; Hanson,
 R.
\newblock {High-fidelity projective read-out of a solid-state spin quantum
 register}.
\newblock {\em Nature} {\bf 2011}, {\em 477},~574--578.

\bibitem[Magesan \em{et~al.}(2015)Magesan, Gambetta, C{\'{o}}rcoles, and
 Chow]{Magesan2015}
Magesan, E.; Gambetta, J.M.; C{\'{o}}rcoles, A.; Chow, J.M.
\newblock {Machine Learning for Discriminating Quantum Measurement Trajectories
 and Improving Readout}.
\newblock {\em Phys. Rev. Lett.} {\bf 2015}, {\em 114},~200501.

\bibitem[Harty \em{et~al.}(2014)Harty, Allcock, Ballance, Guidoni, Janacek,
 Linke, Stacey, and Lucas]{Harty2014}
Harty, T.; Allcock, D.; Ballance, C.; Guidoni, L.; Janacek, H.; Linke, N.;
 Stacey, D.; Lucas, D.
\newblock {High-Fidelity Preparation, Gates, Memory, and Readout of a
 Trapped-Ion Quantum Bit}.
\newblock {\em Phys. Rev. Lett.} {\bf 2014}, {\em 113},~220501.

\bibitem[Shields \em{et~al.}(2015)Shields, Unterreithmeier, de~Leon, Park, and
 Lukin]{Shields2015}
Shields, B.J.; Unterreithmeier, Q.P.; de~Leon, N.P.; Park, H.; Lukin, M.D.
\newblock {Efficient readout of a single spin state in diamond via
 spin-to-charge conversion}.
\newblock {\em Phys. Rev. Lett.} {\bf 2015}, {\em 114},~136402.

\bibitem[Gruber \em{et~al.}(1997)Gruber, Dr{\"{a}}benstedt, Tietz, Fleury,
 Wrachtrup, and von Borczyskowski]{Gruber1997}
Gruber, A.; Dr{\"{a}}benstedt, A.; Tietz, C.; Fleury, L.; Wrachtrup, J.; von
 Borczyskowski, C.
\newblock {Scanning Confocal Optical Microscopy and Magnetic Resonance on
 Single Defect Centers}.
\newblock {\em Science} {\bf 1997}, {\em 276},~2012--2014.

\bibitem[Jelezko \em{et~al.}(2004)Jelezko, Gaebel, Popa, Gruber, and
 Wrachtrup]{Jelezko2004}
Jelezko, F.; Gaebel, T.; Popa, I.; Gruber, A.; Wrachtrup, J.
\newblock {Observation of Coherent Oscillations in a Single Electron Spin}.
\newblock {\em Phys. Rev. Lett.} {\bf 2004}, {\em 92},~76401.

\bibitem[Balasubramanian \em{et~al.}(2009)Balasubramanian, Neumann, Twitchen,
 Markham, Kolesov, Mizuochi, Isoya, Achard, Beck, Tissler, Jacques, Hemmer,
 Jelezko, and Wrachtrup]{Balasubramanian2009}
Balasubramanian, G.; Neumann, P.; Twitchen, D.; Markham, M.; Kolesov, R.;
 Mizuochi, N.; Isoya, J.; Achard,~J.; Beck, J.; Tissler, J.; et al.
\newblock {Ultralong spin coherence time in isotopically engineered diamond}.
\newblock {\em \mbox{Nat. Mater.}} {\bf 2009}, {\em 8},~383.

\bibitem[Dolde \em{et~al.}(2013)Dolde, Jakobi, Naydenov, Zhao, Pezzagna,
 Trautmann, Meijer, Neumann, Jelezko, and Wrachtrup]{Dolde2013}
Dolde, F.; Jakobi, I.; Naydenov, B.; Zhao, N.; Pezzagna, S.; Trautmann, C.;
 Meijer, J.; Neumann, P.; Jelezko, F.; Wrachtrup, J.
\newblock {Room-temperature entanglement between single defect spins in
 diamond}.
\newblock {\em Nat. Phys.} {\bf 2013}, {\em 9},~139--143.

\bibitem[Neumann \em{et~al.}(2008)Neumann, Mizuochi, Rempp, Hemmer, Watanabe,
 Yamasaki, Jacques, Gaebel, Jelezko, and Wrachtrup]{Neumann2008}
Neumann, P.; Mizuochi, N.; Rempp, F.; Hemmer, P.; Watanabe, H.; Yamasaki, S.;
 Jacques, V.; Gaebel, T.; Jelezko, F.; Wrachtrup, J.
\newblock {Multipartite Entanglement Among Single Spins in Diamond}.
\newblock {\em Science} {\bf 2008}, {\em 320},~1326--1329.

\bibitem[Taminiau \em{et~al.}(2014)Taminiau, Cramer, van~der Sar, Dobrovitski,
 and Hanson]{Taminiau2014}
Taminiau, T.H.; Cramer, J.; van~der Sar, T.; Dobrovitski, V.V.; Hanson, R.
\newblock {Universal control and error correction in multi-qubit spin registers
 in diamond}.
\newblock {\em Nat. Nanotechnol.} {\bf 2014}, {\em 9},~171.

\bibitem[van Oort \em{et~al.}(1988)van Oort, Manson, and Glasbeek]{Oort1988}
van Oort, E.; Manson, N.B.; Glasbeek, M.
\newblock {Optically detected spin coherence of the diamond N-V centre in its
 triplet ground state}.
\newblock {\em J. Phys. C Solid State Phys.} {\bf 1988}, {\em 21},~4385.

\bibitem[Goldman \em{et~al.}(2015)Goldman, Doherty, Sipahigil, Yao, Bennett,
 Manson, Kubanek, and Lukin]{Goldman2015a}
Goldman, M.L.; Doherty, M.W.; Sipahigil, A.; Yao, N.Y.; Bennett, S.D.; Manson,
 N.B.; Kubanek, A.; Lukin,~M.D.
\newblock {State-selective intersystem crossing in nitrogen-vacancy centers}.
\newblock {\em Phys. Rev. B} {\bf 2015}, {\em 91},~165201.

\bibitem[Gupta \em{et~al.}(2016)Gupta, Hacquebard, and Childress]{Gupta2016}
Gupta, A.; Hacquebard, L.; Childress, L.
\newblock {Efficient signal processing for time-resolved fluorescence detection
 of nitrogen-vacancy spins in diamond}.
\newblock {\em J. Opt. Soc. Am. B} {\bf 2016}, {\em 33},~B28.

\bibitem[Hopper \em{et~al.}(2016)Hopper, Grote, Exarhos, and
 Bassett]{Hopper2016}
Hopper, D.A.; Grote, R.R.; Exarhos, A.L.; Bassett, L.C.
\newblock {Near-infrared-assisted charge control and spin readout of the
 nitrogen-vacancy center in diamond}.
\newblock {\em Phys. Rev. B} {\bf 2016}, {\em 94},~241201.

\bibitem[Acosta \em{et~al.}(2010)Acosta, Jarmola, Bauch, and
 Budker]{Acosta2010}
Acosta, V.M.; Jarmola, A.; Bauch, E.; Budker, D.
\newblock {Optical properties of the nitrogen-vacancy singlet levels in
 diamond}.
\newblock {\em Phys. Rev. B} {\bf 2010}, {\em 82},~201202.

\bibitem[Waldherr \em{et~al.}(2011)Waldherr, Beck, Steiner, Neumann, Gali,
 Frauenheim, Jelezko, and Wrachtrup]{Waldherr2011a}
Waldherr, G.; Beck, J.; Steiner, M.; Neumann, P.; Gali, A.; Frauenheim, T.;
 Jelezko, F.; Wrachtrup, J.
\newblock {Dark States of Single Nitrogen-Vacancy Centers in Diamond Unraveled
 by Single Shot NMR}.
\newblock {\em Phys. Rev. Lett.} {\bf 2011}, {\em 106},~157601.

\bibitem[Robledo \em{et~al.}(2011)Robledo, Bernien, Van Der~Sar, and
 Hanson]{robledo2011spin}
Robledo, L.; Bernien, H.; Van Der~Sar, T.; Hanson, R.
\newblock Spin dynamics in the optical cycle of single nitrogen-vacancy centres
 in diamond.
\newblock {\em New J. Phys.} {\bf 2011}, {\em 13},~025013.

\bibitem[Plakhotnik \em{et~al.}(1995)Plakhotnik, Moerner, Palm, and
 Wild]{PLAKHOTNIK199583}
Plakhotnik, T.; Moerner, W.; Palm, V.; Wild, U.P.
\newblock Single molecule spectroscopy: maximum emission rate and saturation
 intensity.
\newblock {\em Opt. Commun.} {\bf 1995}, {\em 114},~83--88.

\bibitem[Jamali \em{et~al.}(2014)Jamali, Gerhardt, Rezai, Frenner, Fedder, and
 Wrachtrup]{Jamali2014}
Jamali, M.; Gerhardt, I.; Rezai, M.; Frenner, K.; Fedder, H.; Wrachtrup, J.J.
 {Microscopic diamond solid-immersion-lenses fabricated around single
 defect centers by focused ion beam milling}.
{\em \mbox{Rev. Sci. Instrum.}} {\bf 2014}, {\em 85}, 123703.

\bibitem[Parks \em{et~al.}(2018)Parks, Grote, Hopper, and Bassett]{Parks2018}
Parks, S.M.; Grote, R.R.; Hopper, D.A.; Bassett, L.C.
\newblock {Fabrication of (111)-faced single-crystal diamond plates by laser
 nucleated cleaving}.
\newblock {\em Diam. Rel. Mater.} {\bf 2018}, {\em 84},~20--25.

\bibitem[Miyazaki \em{et~al.}(2014)Miyazaki, Miyamoto, Makino, Kato, Yamasaki,
 Fukui, Doi, Tokuda, Hatano, and Mizuochi]{Miyazaki2014}
Miyazaki, T.; Miyamoto, Y.; Makino, T.; Kato, H.; Yamasaki, S.; Fukui, T.; Doi,
 Y.; Tokuda, N.; Hatano, M.; Mizuochi, N.
\newblock {Atomistic mechanism of perfect alignment of nitrogen-vacancy centers
 in diamond}.
\newblock {\em \mbox{Appl. Phys. Lett.}} {\bf 2014}, {\em 105},~261601.

\bibitem[Lesik \em{et~al.}(2014)Lesik, Tetienne, Tallaire, Achard, Mille,
 Gicquel, Roch, and Jacques]{Lesik2014}
Lesik, M.; Tetienne, J.P.; Tallaire, A.; Achard, J.; Mille, V.; Gicquel, A.;
 Roch, J.F.; Jacques, V.
\newblock {Perfect preferential orientation of nitrogen-vacancy defects in a
 synthetic diamond sample}.
\newblock {\em Appl. Phys. Lett.} {\bf 2014}, {\em 104},~113107.

\bibitem[Michl \em{et~al.}(2014)Michl, Teraji, Zaiser, Jakobi, Waldherr, Dolde,
 Neumann, Doherty, Manson, Isoya, and Wrachtrup]{Michl2014}
Michl, J.; Teraji, T.; Zaiser, S.; Jakobi, I.; Waldherr, G.; Dolde, F.;
 Neumann, P.; Doherty, M.W.; Manson, N.B.; Isoya, J.; Wrachtrup, J.
\newblock {Perfect alignment and preferential orientation of nitrogen-vacancy
 centers during chemical vapor deposition diamond growth on (111) surfaces}.
\newblock {\em Appl. Phys. Lett.} {\bf 2014}, {\em 104},~102407.

\bibitem[Fukui \em{et~al.}(2014)Fukui, Doi, Miyazaki, Miyamoto, Kato,
 Matsumoto, Makino, Yamasaki, Morimoto, Tokuda, Hatano, Sakagawa, Morishita,
 Tashima, Miwa, Suzuki, and Mizuochi]{Fukui2014}
Fukui, T.; Doi, Y.; Miyazaki, T.; Miyamoto, Y.; Kato, H.; Matsumoto, T.;
 Makino, T.; Yamasaki, S.; Morimoto, R.; Tokuda, N.; et al.
\newblock {Perfect selective alignment of nitrogen-vacancy centers in diamond}.
\newblock {\em Appl. Phys. Express} {\bf 2014}, {\em 7},~55201.

\bibitem[Ozawa \em{et~al.}(2017)Ozawa, Tahara, Ishiwata, Hatano, and
 Iwasaki]{Ozawa17}
Ozawa, H.; Tahara, K.; Ishiwata, H.; Hatano, M.; Iwasaki, T.
\newblock {Formation of perfectly aligned nitrogen-vacancy-center ensembles in
 chemical-vapor-deposition-grown diamond (111)}.
\newblock {\em Appl. Phys. Express} {\bf 2017}, {\em 10},~45501.

\bibitem[Hadden \em{et~al.}(2010)Hadden, Harrison, Stanley-Clarke, Marseglia,
 Ho, Patton, O'Brien, and Rarity]{Hadden2010}
Hadden, J.P.; Harrison, J.P.; Stanley-Clarke, A.C.; Marseglia, L.; Ho, Y.L.D.;
 Patton, B.R.; O'Brien, J.L.; Rarity,~J.G.
\newblock {Strongly enhanced photon collection from diamond defect centers
 under microfabricated integrated solid immersion lenses}.
\newblock {\em Appl. Phys. Lett.} {\bf 2010}, {\em 97},~241901.

\bibitem[Marseglia \em{et~al.}(2011)Marseglia, Hadden, Stanley-Clarke,
 Harrison, Patton, Ho, Naydenov, Jelezko, Meijer, Dolan, Smith, Rarity, and
 O'brien]{Marseglia2011}
Marseglia, L.; Hadden, J.P.; Stanley-Clarke, A.C.; Harrison, J.P.; Patton, B.;
 Ho, Y.L.D.; Naydenov, B.; Jelezko,~F.; Meijer, J.; Dolan, P.R.; et al.
\newblock {Nanofabricated solid immersion lenses registered to single emitters
 in diamond}.
\newblock {\em Appl. Phys. Lett.} {\bf 2011}, {\em 98},~14--17.

\bibitem[Grote \em{et~al.}(2017)Grote, Huang, Mann, Hopper, Exarhos, Lopez,
 Garnett, and Bassett]{Grote2017}
Grote, R.R.; Huang, T.Y.; Mann, S.A.; Hopper, D.A.; Exarhos, A.L.; Lopez, G.G.;
 Garnett, E.C.; Bassett, L.C.
\newblock {Imaging a Nitrogen-Vacancy Center with a Diamond Immersion Metalens.} \emph{arXiv}
 {\bf 2017}, arxiv:1711.00901.

\bibitem[Neu \em{et~al.}(2014)Neu, Appel, Ganzhorn, Miguel-Sánchez, Lesik,
 Mille, Jacques, Tallaire, Achard, and Maletinsky]{Neu2014}
Neu, E.; Appel, P.; Ganzhorn, M.; Miguel-Sánchez, J.; Lesik, M.; Mille, V.;
 Jacques, V.; Tallaire, A.; Achard, J.; Maletinsky, P.
\newblock Photonic nano-structures on (111)-oriented diamond.
\newblock {\em Appl. Phys. Lett.} {\bf 2014}, {\em 104},~153108.

\bibitem[Riedel \em{et~al.}(2017)Riedel, S\"ollner, Shields, Starosielec,
 Appel, Neu, Maletinsky, and Warburton]{Riedel2017}
Riedel, D.; S\"ollner, I.; Shields, B.J.; Starosielec, S.; Appel, P.; Neu, E.;
 Maletinsky, P.; Warburton, R.J.
\newblock Deterministic Enhancement of Coherent Photon Generation from a
 Nitrogen-Vacancy Center in Ultrapure Diamond.
\newblock {\em Phys. Rev. X} {\bf 2017}, {\em 7},~031040.

\bibitem[Mouradian \em{et~al.}(2015)Mouradian, Schr{\"{o}}der, Poitras, Li,
 Goldstein, Chen, Walsh, Cardenas, Markham, Twitchen, Lipson, and
 Englund]{Mouradian2015}
Mouradian, S.L.; Schr{\"{o}}der, T.; Poitras, C.B.; Li, L.; Goldstein, J.;
 Chen, E.H.; Walsh, M.; Cardenas, J.; Markham,~M.L.; Twitchen, D.J.; et al.
\newblock {Scalable Integration of Long-Lived Quantum Memories into a Photonic
 Circuit}.
\newblock {\em Phys. Rev. X} {\bf 2015}, {\em 5},~31009.

\bibitem[Babinec \em{et~al.}(2010)Babinec, Hausmann, Khan, Zhang, Maze, Hemmer,
 and Lon{\v{c}}ar]{Babinec2010}
Babinec, T.M.; Hausmann, B.J.M.; Khan, M.; Zhang, Y.; Maze, J.R.; Hemmer, P.R.;
 Lon{\v{c}}ar, M.
\newblock {A diamond nanowire single-photon source}.
\newblock {\em Nat. Nanotechnol.} {\bf 2010}, {\em 5},~195.

\bibitem[Momenzadeh \em{et~al.}(2015)Momenzadeh, St{\"{o}}hr, de~Oliveira,
 Brunner, Denisenko, Yang, Reinhard, and Wrachtrup]{Momenzadeh2015}
Momenzadeh, S.A.; St{\"{o}}hr, R.J.; de~Oliveira, F.F.; Brunner, A.; Denisenko,
 A.; Yang, S.; Reinhard, F.; Wrachtrup,~J.
\newblock {Nanoengineered Diamond Waveguide as a Robust Bright Platform for
 Nanomagnetometry Using Shallow Nitrogen Vacancy Centers}.
\newblock {\em Nano Lett.} {\bf 2015}, {\em 15},~165--169.

\bibitem[Maletinsky \em{et~al.}(2012)Maletinsky, Hong, Grinolds, Hausmann,
 Lukin, Walsworth, Loncar, and Yacoby]{Maletinsky2012}
Maletinsky, P.; Hong, S.; Grinolds, M.S.; Hausmann, B.; Lukin, M.D.; Walsworth,
 R.L.; Loncar, M.; Yacoby,~A.
\newblock {A robust scanning diamond sensor for nanoscale imaging with single
 nitrogen-vacancy centres}.
\newblock {\em \mbox{Nat. Nanotechnol.}} {\bf 2012}, {\em 7},~320--324.

\bibitem[Appel \em{et~al.}(2016)Appel, Neu, Ganzhorn, Barfuss, Batzer, Gratz,
 Tschöpe, and Maletinsky]{Appel2016}
Appel, P.; Neu, E.; Ganzhorn, M.; Barfuss, A.; Batzer, M.; Gratz, M.; Tschöpe,
 A.; Maletinsky, P.
\newblock Fabrication of all diamond scanning probes for nanoscale
 magnetometry.
\newblock {\em Rev. Sci. Instrum.} {\bf 2016}, {\em 87},~063703.

\bibitem[Hausmann \em{et~al.}(2012)Hausmann, Shields, Quan, Maletinsky,
 McCutcheon, Choy, Babinec, Kubanek, Yacoby, Lukin, and Loncar]{Hausmann2012}
Hausmann, B.J.M.; Shields, B.; Quan, Q.; Maletinsky, P.; McCutcheon, M.; Choy,
 J.T.; Babinec, T.M.; Kubanek,~A.; Yacoby, A.; Lukin, M.D.; et al.
\newblock {Integrated Diamond Networks for Quantum Nanophotonics}.
\newblock {\em \mbox{Nano Lett.}} {\bf 2012}, {\em 12},~1578--1582.

\bibitem[Gould \em{et~al.}(2016)Gould, Chakravarthi, Christen, Thomas,
 Dadgostar, Song, Lee, Hatami, and Fu]{Gould2016}
Gould, M.; Chakravarthi, S.; Christen, I.R.; Thomas, N.; Dadgostar, S.; Song,
 Y.; Lee, M.L.; Hatami, F.; Fu, K.M.C.
\newblock {Large-scale GaP-on-diamond integrated photonics platform for NV
 center-based quantum information}.
\newblock {\em J. Opt. Soc. Am. B} {\bf 2016}, {\em 33},~B35--B42.

\bibitem[Faraon \em{et~al.}(2011)Faraon, Barclay, Santori, Fu, and
 Beausoleil]{faraon2011resonant}
Faraon, A.; Barclay, P.E.; Santori, C.; Fu, K.M.C.; Beausoleil, R.G.
\newblock Resonant enhancement of the zero-phonon emission from a colour centre
 in a diamond cavity.
\newblock {\em Nat. Photonics} {\bf 2011}, {\em 5},~301.

\bibitem[Hausmann \em{et~al.}(2013)Hausmann, Shields, Quan, Chu, de~Leon,
 Evans, Burek, Zibrov, Markham, Twitchen, Park, Lukin, and
 Loncar]{Hausmann2013}
Hausmann, B.J.M.; Shields, B.J.; Quan, Q.; Chu, Y.; de~Leon, N.P.; Evans, R.;
 Burek, M.J.; Zibrov, A.S.; Markham, M.; Twitchen, D.J.; et al.
\newblock {Coupling of NV Centers to Photonic Crystal Nanobeams in Diamond}.
\newblock {\em Nano Lett.} {\bf 2013}, {\em 13},~5791--5796.

\bibitem[Faraon \em{et~al.}(2012)Faraon, Santori, Huang, Acosta, and
 Beausoleil]{Faraon2012}
Faraon, A.; Santori, C.; Huang, Z.; Acosta, V.M.; Beausoleil, R.G.
\newblock {Coupling of Nitrogen-Vacancy Centers to Photonic Crystal Cavities in
 Monocrystalline Diamond}.
\newblock {\em Phys. Rev. Lett.} {\bf 2012}, {\em 109},~33604.

\bibitem[Grote and Bassett(2016)]{Grote2016}
Grote, R.R.; Bassett, L.C.
\newblock {Single-mode optical waveguides on native high-refractive-index
 substrates}.
\newblock {\em \mbox{APL Photonics}} {\bf 2016}, {\em 1},~71302.

\bibitem[Mouradian and Englund(2017)]{mouradian2017tunable}
Mouradian, S.L.; Englund, D.
\newblock A tunable waveguide-coupled cavity design for scalable interfaces to
 solid-state quantum emitters.
\newblock {\em \mbox{APL Photonics}} {\bf 2017}, {\em 2},~046103.

\bibitem[Johnson \em{et~al.}(2015)Johnson, Dolan, Grange, Trichet, Hornecker,
 Chen, Weng, Hughes, Watt, Auff{\`{e}}ves, and Smith]{Johnson2015}
Johnson, S.; Dolan, P.R.; Grange, T.; Trichet, A.A.P.; Hornecker, G.; Chen,
 Y.C.; Weng, L.; Hughes, G.M.; Watt,~A.A.R.; Auff{\`{e}}ves, A.; et al.
\newblock {Tunable cavity coupling of the zero phonon line of a
 nitrogen-vacancy defect in diamond}.
\newblock {\em New J. Phys.} {\bf 2015}, {\em 17},~122003.

\bibitem[Bogdanovi{\'{c}} \em{et~al.}(2017)Bogdanovi{\'{c}}, Liddy, van Dam,
 Coenen, Fink, Lon{\v{c}}ar, and Hanson]{Bogdanovic2017}
Bogdanovi{\'{c}}, S.; Liddy, M.S.Z.; van Dam, S.B.; Coenen, L.C.; Fink, T.;
 Lon{\v{c}}ar, M.; Hanson, R.
\newblock {Robust nano-fabrication of an integrated platform for spin control
 in a tunable microcavity}.
\newblock {\em APL Photonics} {\bf 2017}, {\em 2},~126101.

\bibitem[Purcell(1946)]{Purcell1946}
Purcell, E.M.
\newblock \emph{Phys. Rev.} {\bf 1946}, {\em 69},~681.

\bibitem[Vahala(2003)]{Vahala2003}
Vahala, K.J.
\newblock {Optical microcavities}.
\newblock {\em Nature} {\bf 2003}, {\em 424},~839.

\bibitem[Wolf \em{et~al.}(2015)Wolf, Rosenberg, Rapaport, and
 Bar-Gill]{Wolf2015}
Wolf, S.A.; Rosenberg, I.; Rapaport, R.; Bar-Gill, N.
\newblock {Purcell-enhanced optical spin readout of nitrogen-vacancy centers in
 diamond}.
\newblock {\em Phys. Rev. B} {\bf 2015}, {\em 92},~235410.

\bibitem[Wang \em{et~al.}(2007)Wang, Hanson, Awschalom, Hu, Feygelson, Yang,
 and Butler]{Wang2007}
Wang, C.F.; Hanson, R.; Awschalom, D.D.; Hu, E.L.; Feygelson, T.; Yang, J.;
 Butler, J.E.
\newblock {Fabrication and characterization of two-dimensional photonic crystal
 microcavities in nanocrystalline diamond}.
\newblock {\em \mbox{Appl. Phys. Lett.}} {\bf 2007}, {\em 91},~201112.

\bibitem[Wolters \em{et~al.}(2010)Wolters, Schell, Kewes, N{\"{u}}sse,
 Schoengen, D{\"{o}}scher, Hannappel, L{\"{o}}chel, Barth, and
 Benson]{Wolters2010}
Wolters, J.; Schell, A.W.; Kewes, G.; N{\"{u}}sse, N.; Schoengen, M.;
 D{\"{o}}scher, H.; Hannappel, T.; L{\"{o}}chel, B.; Barth,~M.; Benson, O.
\newblock {Enhancement of the zero phonon line emission from a single nitrogen
 vacancy center in a nanodiamond via coupling to a photonic crystal cavity}.
\newblock {\em Appl. Phys. Lett.} {\bf 2010}, {\em 97},~141108.

\bibitem[Barclay \em{et~al.}(2009)Barclay, Santori, Fu, Beausoleil, and
 Painter]{barclay2009coherent}
Barclay, P.E.; Santori, C.; Fu, K.M.; Beausoleil, R.G.; Painter, O.
\newblock Coherent interference effects in a nano-assembled diamond NV center
 cavity-QED system.
\newblock {\em Opt. Express} {\bf 2009}, {\em 17},~8081--8197.

\bibitem[Lee \em{et~al.}(2014)Lee, Bracher, Cui, Ohno, McLellan, Zhang,
 Andrich, Alem\'an, Russell, Magyar, Aharonovich, Bleszynski~Jayich,
 Awschalom, and Hu]{Lee2014}
Lee, J.C.; Bracher, D.O.; Cui, S.; Ohno, K.; McLellan, C.A.; Zhang, X.;
 Andrich, P.; Alem\'an, B.; Russell, K.J.; Magyar, A.P.; et al.
\newblock Deterministic coupling of delta-doped nitrogen vacancy centers to a
 nanobeam photonic crystal cavity.
\newblock {\em Appl. Phys. Lett.} {\bf 2014}, {\em 105},~261101.

\bibitem[Englund \em{et~al.}(2010)Englund, Shields, Rivoire, Hatami,
 Vu{\v{c}}kovi{\'{c}}, Park, and Lukin]{Englund2010}
Englund, D.; Shields, B.; Rivoire, K.; Hatami, F.; Vu{\v{c}}kovi{\'{c}}, J.;
 Park, H.; Lukin, M.D.
\newblock {Deterministic Coupling of a Single Nitrogen Vacancy Center to a
 Photonic Crystal Cavity}.
\newblock {\em Nano Lett.} {\bf 2010}, {\em 10},~3922--3926.


\bibitem[K{\"{u}}hn \em{et~al.}(2006)K{\"{u}}hn, H{\aa}kanson, Rogobete, and
 Sandoghdar]{Kuhn2006}
K{\"{u}}hn, S.; H{\aa}kanson, U.; Rogobete, L.; Sandoghdar, V.
\newblock {Enhancement of Single-Molecule Fluorescence Using a Gold
 Nanoparticle as an Optical Nanoantenna}.
\newblock {\em Phys. Rev. Lett.} {\bf 2006}, {\em 97},~17402.

\bibitem[Akimov \em{et~al.}(2007)Akimov, Mukherjee, Yu, Chang, Zibrov, Hemmer,
 Park, and Lukin]{Akimov2007}
Akimov, A.V.; Mukherjee, A.; Yu, C.L.; Chang, D.E.; Zibrov, A.S.; Hemmer, P.R.;
 Park, H.; Lukin, M.D.
\newblock {Generation of single optical plasmons in metallic nanowires coupled
 to quantum dots}.
\newblock {\em Nature} {\bf 2007}, {\em 450},~402.

\bibitem[Schietinger \em{et~al.}(2009)Schietinger, Barth, Aichele, and
 Benson]{Schietinger2009}
Schietinger, S.; Barth, M.; Aichele, T.; Benson, O.
\newblock {Plasmon-Enhanced Single Photon Emission from a Nanoassembled
 Metal-Diamond Hybrid Structure at Room Temperature}.
\newblock {\em Nano Lett.} {\bf 2009}, {\em 9},~1694--1698.

\bibitem[Akselrod \em{et~al.}(2014)Akselrod, Argyropoulos, Hoang, Cirac{\`\i},
 Fang, Huang, Smith, and Mikkelsen]{akselrod2014probing}
Akselrod, G.M.; Argyropoulos, C.; Hoang, T.B.; Cirac{\`\i}, C.; Fang, C.;
 Huang, J.; Smith, D.R.; Mikkelsen, M.H.
\newblock Probing the mechanisms of large Purcell enhancement in plasmonic
 nanoantennas.
\newblock {\em Nat. Photonics} {\bf 2014}, {\em 8},~835.

\bibitem[Hoang \em{et~al.}(2015)Hoang, Akselrod, and
 Mikkelsen]{hoang2015ultrafast}
Hoang, T.B.; Akselrod, G.M.; Mikkelsen, M.H.
\newblock Ultrafast room-temperature single photon emission from quantum dots
 coupled to plasmonic nanocavities.
\newblock {\em Nano Lett.} {\bf 2015}, {\em 16},~270--275.

\bibitem[Anger \em{et~al.}(2006)Anger, Bharadwaj, and Novotny]{Anger2006}
Anger, P.; Bharadwaj, P.; Novotny, L.
\newblock {Enhancement and Quenching of Single-Molecule Fluorescence}.
\newblock {\em \mbox{Phys. Rev. Lett.}} {\bf 2006}, {\em 96},~113002.

\bibitem[Shalaginov \em{et~al.}(2013)Shalaginov, Ishii, Liu, Liu, Irudayaraj,
 Lagutchev, Kildishev, and Shalaev]{Shalaginov2013}
Shalaginov, M.Y.; Ishii, S.; Liu, J.; Liu, J.; Irudayaraj, J.; Lagutchev, A.;
 Kildishev, A.V.; Shalaev, V.M.
\newblock {Broadband enhancement of spontaneous emission from nitrogen-vacancy
 centers in nanodiamonds by hyperbolic metamaterials}.
\newblock {\em Appl. Phys. Lett.} {\bf 2013}, {\em 102},~173114.

\bibitem[Bulu \em{et~al.}(2011)Bulu, Babinec, Hausmann, Choy, and
 Loncar]{Bulu2011}
Bulu, I.; Babinec, T.; Hausmann, B.; Choy, J.T.; Loncar, M.
\newblock {Plasmonic resonators for enhanced diamond NV- center single photon
 sources}.
\newblock {\em Opt. Express} {\bf 2011}, {\em 19},~5268--5276.

\bibitem[Riedel \em{et~al.}(2014)Riedel, Rohner, Ganzhorn, Kaldewey, Appel,
 Neu, Warburton, and Maletinsky]{Riedel2014}
Riedel, D.; Rohner, D.; Ganzhorn, M.; Kaldewey, T.; Appel, P.; Neu, E.;
 Warburton, R.; Maletinsky, P.
\newblock {Low-Loss Broadband Antenna for Efficient Photon Collection from a
 Coherent Spin in Diamond}.
\newblock {\em \mbox{Phys. Rev. Appl.}} {\bf 2014}, {\em 2},~64011.

\bibitem[Karamlou \em{et~al.}(2018)Karamlou, Trusheim, and
 Englund]{Karamlou2018}
Karamlou, A.; Trusheim, M.E.; Englund, D.
\newblock {Metal-dielectric antennas for efficient photon collection from
 diamond color centers}.
\newblock {\em Opt. Express} {\bf 2018}, {\em 26},~3341--3352.

\bibitem[Bogdanov \em{et~al.}(2017)Bogdanov, Shalaginov, Akimov, Lagutchev,
 Kapitanova, Liu, Woods, Ferrera, Belov, Irudayaraj, Boltasseva, and
 Shalaev]{Bogdanov2017}
Bogdanov, S.; Shalaginov, M.Y.; Akimov, A.; Lagutchev, A.S.; Kapitanova, P.;
 Liu, J.; Woods, D.; Ferrera, M.; Belov, P.; Irudayaraj, J.; et al.
\newblock {Electron spin contrast of Purcell-enhanced nitrogen-vacancy
 ensembles in nanodiamonds}.
\newblock {\em Phys. Rev. B} {\bf 2017}, {\em 96},~035146.

\bibitem[Babinec \em{et~al.}(2012)Babinec, Fedder, Choy, Bulu, Doherty, Hemmer,
 Wrachtrup, and Loncar]{Babinec2012}
Babinec, T.M.; Fedder, H.; Choy, J.; Bulu, I.; Doherty, M.; Hemmer, P.;
 Wrachtrup, J.; Loncar, M.
\newblock {Design of Diamond Photonic Devices for Spintronics}.
\newblock In Proceedings of the 2012 Conference on Lasers and Electro-Optics (CLEO), San Jose, CA, USA, 6--11 May 2012; p. JW1I.6.

\bibitem[Hopper \em{et~al.}(2018)Hopper, Grote, Parks, and Bassett]{Hopper2018}
Hopper, D.A.; Grote, R.R.; Parks, S.M.; Bassett, L.C.
\newblock {Amplified Sensitivity of Nitrogen-Vacancy Spins in Nanodiamonds
 Using All-Optical Charge Readout}.
\newblock {\em ACS Nano} {\bf 2018}, \emph{12}, 4678--4686.

\bibitem[Tamarat \em{et~al.}(2008)Tamarat, Manson, Harrison, McMurtrie,
 Nizovtsev, Santori, Neumann, Beausoleil, Gaebel, Jelezko, Hemmer, and
 Wrachtrup]{Tamarat2008}
Tamarat, P.; Manson, N.B.; Harrison, J.P.; McMurtrie, R.L.; Nizovtsev, A.;
 Santori, C.; Neumann, P.; Beausoleil,~R.G.; Gaebel, T.; Jelezko, F.; et al.
\newblock {Spin-flip and spin-conserving optical transitions of the
 nitrogen-vacancy centre in diamond}.
\newblock {\em New J. Phys.} {\bf 2008}, {\em 10},~45004.

\bibitem[Batalov \em{et~al.}(2009)Batalov, Jacques, Kaiser, Siyushev, Neumann,
 Rogers, McMurtrie, Manson, Jelezko, and Wrachtrup]{batalov2009low}
Batalov, A.; Jacques, V.; Kaiser, F.; Siyushev, P.; Neumann, P.; Rogers, L.;
 McMurtrie, R.; Manson, N.; Jelezko, F.; Wrachtrup, J.
\newblock Low temperature studies of the excited-state structure of negatively
 charged nitrogen-vacancy color centers in diamond.
\newblock {\em Phys. Rev. Lett.} {\bf 2009}, {\em 102},~195506.

\bibitem[Fu \em{et~al.}(2009)Fu, Santori, Barclay, Rogers, Manson, and
 Beausoleil]{fu2009observation}
Fu, K.M.C.; Santori, C.; Barclay, P.E.; Rogers, L.J.; Manson, N.B.; Beausoleil,
 R.G.
\newblock Observation of the dynamic Jahn-Teller effect in the excited states
 of nitrogen-vacancy centers in diamond.
\newblock {\em Phys. Rev. Lett.} {\bf 2009}, {\em 103},~256404.

\bibitem[Fuchs \em{et~al.}(2010)Fuchs, Dobrovitski, Toyli, Heremans, Weis,
 Schenkel, and Awschalom]{fuchs2010excited}
Fuchs, G.; Dobrovitski, V.; Toyli, D.; Heremans, F.; Weis, C.; Schenkel, T.;
 Awschalom, D.
\newblock Excited-state spin coherence of a single nitrogen--vacancy centre in
 diamond.
\newblock {\em Nat. Phys.} {\bf 2010}, {\em 6},~668.

\bibitem[Buckley \em{et~al.}(2010)Buckley, Fuchs, Bassett, and
 Awschalom]{Buckley2010}
Buckley, B.B.; Fuchs, G.D.; Bassett, L.C.; Awschalom, D.D.
\newblock {Spin-Light Coherence for Single-Spin Measurement and Control in
 Diamond}.
\newblock {\em Science} {\bf 2010}, {\em 330},~1212--1215.

\bibitem[Togan \em{et~al.}(2010)Togan, Chu, Trifonov, Jiang, Maze, Childress,
 Dutt, S{\o}rensen, Hemmer, Zibrov, et~al.]{togan2010quantum}
Togan, E.; Chu, Y.; Trifonov, A.; Jiang, L.; Maze, J.; Childress, L.; Dutt,
 M.G.; S{\o}rensen, A.S.; Hemmer, P.; Zibrov, A.S.; et al.
\newblock Quantum entanglement between an optical photon and a solid-state spin
 qubit.
\newblock {\em Nature} {\bf 2010}, {\em 466},~730.

\bibitem[Yale \em{et~al.}(2013)Yale, Buckley, Christle, Burkard, Heremans,
 Bassett, and Awschalom]{Yale2013}
Yale, C.G.; Buckley, B.B.; Christle, D.J.; Burkard, G.; Heremans, F.J.;
 Bassett, L.C.; Awschalom, D.D.
\newblock {All-optical control of a solid-state spin using coherent dark
 states}.
\newblock {\em Proc. Natl. Acad. Sci. USA} {\bf 2013}, {\em 110},~7595--7600.

\bibitem[Bassett \em{et~al.}(2014)Bassett, Heremans, Christle, Yale, Burkard,
 Buckley, and Awschalom]{Bassett2014}
Bassett, L.C.; Heremans, F.J.; Christle, D.J.; Yale, C.G.; Burkard, G.;
 Buckley, B.B.; Awschalom, D.D.
\newblock {Ultrafast optical control of orbital and spin dynamics in a
 solid-state defect}.
\newblock {\em Science} {\bf 2014}, {\em 345},~1333--1337.

\bibitem[Olmschenk \em{et~al.}(2007)Olmschenk, Younge, Moehring, Matsukevich,
 Maunz, and Monroe]{Olmschenk2007}
Olmschenk, S.; Younge, K.C.; Moehring, D.L.; Matsukevich, D.N.; Maunz, P.;
 Monroe, C.
\newblock {Manipulation and detection of a trapped $\textrm{Yb}^+$ hyperfine
 qubit}.
\newblock {\em Phys. Rev. A} {\bf 2007}, {\em 76},~52314.

\bibitem[Vamivakas \em{et~al.}(2010)Vamivakas, Lu, Matthiesen, Zhao,
 F{\"{a}}lt, Badolato, and Atat{\"{u}}re]{Vamivakas2010}
Vamivakas, A.N.; Lu, C.Y.; Matthiesen, C.; Zhao, Y.; F{\"{a}}lt, S.; Badolato,
 A.; Atat{\"{u}}re, M.
\newblock {Observation of spin-dependent quantum jumps via quantum dot
 resonance fluorescence}.
\newblock {\em Nature} {\bf 2010}, {\em 467},~297.

\bibitem[Bassett \em{et~al.}(2011)Bassett, Heremans, Yale, Buckley, and
 Awschalom]{bassett2011electrical}
Bassett, L.C.; Heremans, F.J.; Yale, C.; Buckley, B.; Awschalom, D.D.
\newblock Electrical tuning of single nitrogen-vacancy center optical
 transitions enhanced by photoinduced fields.
\newblock {\em Phys. Rev. Lett.} {\bf 2011}, {\em 107},~266403.

\bibitem[Blok \em{et~al.}(2014)Blok, Bonato, Markham, Twitchen, Dobrovitski,
 and Hanson]{Blok2014}
Blok, M.S.; Bonato, C.; Markham, M.L.; Twitchen, D.J.; Dobrovitski, V.V.;
 Hanson, R.
\newblock {Manipulating a qubit through the backaction of sequential partial
 measurements and real-time feedback}.
\newblock {\em Nat. Phys.} {\bf 2014}, {\em 10},~189.

\bibitem[Bernien \em{et~al.}(2013)Bernien, Hensen, Pfaff, Koolstra, Blok,
 Robledo, Taminiau, Markham, Twitchen, Childress, and Hanson]{Bernien2013}
Bernien, H.; Hensen, B.; Pfaff, W.; Koolstra, G.; Blok, M.S.; Robledo, L.;
 Taminiau, T.H.; Markham, M.; Twitchen, D.J.; Childress, L.; et al.
\newblock {Heralded entanglement between solid-state qubits separated by three
 metres}.
\newblock {\em Nature} {\bf 2013}, {\em 497},~86--90.

\bibitem[Cramer \em{et~al.}(2016)Cramer, Kalb, Rol, Hensen, Blok, Markham,
 Twitchen, Hanson, and Taminiau]{Cramer2016}
Cramer, J.; Kalb, N.; Rol, M.A.; Hensen, B.; Blok, M.S.; Markham, M.; Twitchen,
 D.J.; Hanson, R.; Taminiau, T.H.
\newblock {Repeated quantum error correction on a continuously encoded qubit by
 real-time feedback}.
\newblock {\em \mbox{Nat. Commun.}} {\bf 2016}, {\em 7},~11526.

\bibitem[Terblanche \em{et~al.}(2001)Terblanche, Reynhardt, and van
 Wyk]{Terblanche2001}
Terblanche, C.J.; Reynhardt, E.C.; van Wyk, J.A.
\newblock {13C Spin–Lattice Relaxation in Natural Diamond: Zeeman Relaxation
 at 4.7 T and 300 K Due to Fixed Paramagnetic Nitrogen Defects}.
\newblock {\em Solid State Nucl. Magn. Reson.} {\bf 2001}, {\em 20},~1--22.

\bibitem[Reiserer \em{et~al.}(2016)Reiserer, Kalb, Blok, van Bemmelen,
 Taminiau, Hanson, Twitchen, and Markham]{reiserer2016robust}
Reiserer, A.; Kalb, N.; Blok, M.S.; van Bemmelen, K.J.; Taminiau, T.H.; Hanson,
 R.; Twitchen, D.J.; Markham,~M.
\newblock Robust quantum-network memory using decoherence-protected subspaces
 of nuclear spins.
\newblock {\em \mbox{Phys. Rev. X}} {\bf 2016}, {\em 6},~021040.

\bibitem[Kalb \em{et~al.}(2017)Kalb, Reiserer, Humphreys, Bakermans, Kamerling,
 Nickerson, Benjamin, Twitchen, Markham, and Hanson]{kalb2017entanglement}
Kalb, N.; Reiserer, A.A.; Humphreys, P.C.; Bakermans, J.J.; Kamerling, S.J.;
 Nickerson, N.H.; Benjamin, S.C.; Twitchen, D.J.; Markham, M.; Hanson, R.
\newblock Entanglement distillation between solid-state quantum network nodes.
\newblock {\em Science} {\bf 2017}, {\em 356},~928--932.

\bibitem[Jiang \em{et~al.}(2009)Jiang, Hodges, Maze, Maurer, Taylor, Cory,
 Hemmer, Walsworth, Yacoby, Zibrov, and Lukin]{Jiang2009}
Jiang, L.; Hodges, J.S.; Maze, J.R.; Maurer, P.; Taylor, J.M.; Cory, D.G.;
 Hemmer, P.R.; Walsworth, R.L.; Yacoby,~A.; Zibrov, A.S.; et al.
\newblock {Repetitive Readout of a Single Electronic Spin via Quantum Logic
 with Nuclear Spin Ancillae}.
\newblock {\em Science} {\bf 2009}, {\em 326},~267--272.

\bibitem[Steiner \em{et~al.}(2010)Steiner, Neumann, Beck, Jelezko, and
 Wrachtrup]{Steiner2010}
Steiner, M.; Neumann, P.; Beck, J.; Jelezko, F.; Wrachtrup, J.
\newblock {Universal enhancement of the optical readout fidelity of single
 electron spins at nitrogen-vacancy centers in diamond}.
\newblock {\em Phys. Rev. B} {\bf 2010}, {\em 81},~35205.

\bibitem[Dr{\'{e}}au \em{et~al.}(2013)Dr{\'{e}}au, Spinicelli, Maze, Roch, and
 Jacques]{Dreau2013}
Dr{\'{e}}au, A.; Spinicelli, P.; Maze, J.R.; Roch, J.F.; Jacques, V.
\newblock {Single-Shot Readout of Multiple Nuclear Spin Qubits in Diamond under
 Ambient Conditions}.
\newblock {\em Phys. Rev. Lett.} {\bf 2013}, {\em 110},~60502.

\bibitem[Neumann \em{et~al.}(2009)Neumann, Kolesov, Jacques, Beck, Tisler,
 Batalov, Rogers, Manson, Balasubramanian, Jelezko, and
 Wrachtrup]{Neumann2009}
Neumann, P.; Kolesov, R.; Jacques, V.; Beck, J.; Tisler, J.; Batalov, A.;
 Rogers, L.; Manson, N.B.; Balasubramanian, G.; Jelezko, F.; et al.
\newblock {Excited-state spectroscopy of single NV defects in diamond using
 optically detected magnetic resonance}.
\newblock {\em New J. Phys.} {\bf 2009}, {\em 11},~13017.

\bibitem[D'Anjou and Coish(2017)]{DAnjou2017}
D'Anjou, B.; Coish, W.A.
\newblock {Enhancing qubit readout through dissipative sub-Poissonian
 dynamics}.
\newblock {\em \mbox{Phys. Rev. A}} {\bf 2017}, {\em 96},~52321.

\bibitem[H{\"{a}}berle \em{et~al.}(2017)H{\"{a}}berle, Oeckinghaus,
 Schmid-Lorch, Pfender, de~Oliveira, Momenzadeh, Finkler, and
 Wrachtrup]{Haberle2017}
H{\"{a}}berle, T.; Oeckinghaus, T.; Schmid-Lorch, D.; Pfender, M.; de~Oliveira,
 F.F.; Momenzadeh, S.A.; Finkler, A.; Wrachtrup, J.
\newblock {Nuclear quantum-assisted magnetometer}.
\newblock {\em Rev. Sci. Instrum.} {\bf 2017}, {\em 88},~13702.

\bibitem[Waldherr \em{et~al.}(2012)Waldherr, Beck, Neumann, Said, Nitsche,
 Markham, Twitchen, Twamley, Jelezko, and Wrachtrup]{Waldherr2012}
Waldherr, G.; Beck, J.; Neumann, P.; Said, R.S.; Nitsche, M.; Markham, M.L.;
 Twitchen, D.J.; Twamley, J.; Jelezko, F.; Wrachtrup, J.
\newblock {High-dynamic-range magnetometry with a single nuclear spin in
 diamond}.
\newblock {\em \mbox{Nat. Nanotechnol.}} {\bf 2012}, {\em 7},~105--108.

\bibitem[Zaiser \em{et~al.}(2016)Zaiser, Rendler, Jakobi, Wolf, Lee, Wagner,
 Neumann, Bergholm, Schulte-Herbr{\"{u}}ggen, Neumann, and
 Wrachtrup]{Zaiser2016}
Zaiser, S.; Rendler, T.; Jakobi, I.; Wolf, T.; Lee, S.y.; Wagner, S.; Neumann,
 P.; Bergholm, V.; Schulte-Herbr{\"{u}}ggen, T.; Neumann, P.; et al.
\newblock {Enhancing quantum sensing sensitivity by a quantum memory}.
\newblock {\em Nat. Commun.} {\bf 2016}, {\em 7},~12279.

\bibitem[Pfender \em{et~al.}(2017)Pfender, Aslam, Sumiya, Onoda, Neumann,
 Isoya, Meriles, and Wrachtrup]{Pfender2017}
Pfender, M.; Aslam, N.; Sumiya, H.; Onoda, S.; Neumann, P.; Isoya, J.; Meriles,
 C.A.; Wrachtrup, J.
\newblock {Nonvolatile nuclear spin memory enables sensor-unlimited nanoscale
 spectroscopy of small spin clusters}.
\newblock {\em Nat. Commun.} {\bf 2017}, {\em 8},~834.

\bibitem[Elzerman \em{et~al.}(2004)Elzerman, Hanson, {Willems van Beveren},
 Witkamp, Vandersypen, and Kouwenhoven]{Elzerman2004}
Elzerman, J.M.; Hanson, R.; {Willems van Beveren}, L.H.; Witkamp, B.;
 Vandersypen, L.M.K.; Kouwenhoven,~L.P.
 {Single-shot read-out of an individual electron spin in a quantum
 dot}.
 {\em Nature} {\bf 2004}, {\em 430},~431--435.

\bibitem[Morello \em{et~al.}(2010)Morello, Pla, Zwanenburg, Chan, Tan, Huebl,
 Mottonen, Nugroho, Yang, van Donkelaar, Alves, Jamieson, Escott, Hollenberg,
 Clark, and Dzurak]{Morello2010}
Morello, A.; Pla, J.J.; Zwanenburg, F.A.; Chan, K.W.; Tan, K.Y.; Huebl, H.;
 Mottonen, M.; Nugroho, C.D.; Yang,~C.; van Donkelaar, J.A.; et al.
\newblock {Single-shot readout of an electron spin in silicon}.
\newblock {\em Nature} {\bf 2010}, {\em 467},~687--691.

\bibitem[Waldherr \em{et~al.}(2011)Waldherr, Neumann, Huelga, Jelezko, and
 Wrachtrup]{Waldherr2011}
Waldherr, G.; Neumann, P.; Huelga, S.F.; Jelezko, F.; Wrachtrup, J.
\newblock {Violation of a Temporal Bell Inequality for Single Spins in a
 Diamond Defect Center}.
\newblock {\em Phys. Rev. Lett.} {\bf 2011}, {\em 107},~090401.

\bibitem[Jayakumar \em{et~al.}(2016)Jayakumar, Henshaw, Dhomkar, Pagliero,
 Laraoui, Manson, Albu, Doherty, and Meriles]{Jayakumar2016}
Jayakumar, H.; Henshaw, J.; Dhomkar, S.; Pagliero, D.; Laraoui, A.; Manson,
 N.B.; Albu, R.; Doherty, M.W.; Meriles, C.A.
 {Optical patterning of trapped charge in nitrogen-doped diamond}.
 {\em Nat. Commun.} {\bf 2016}, {\em 7}, 12660.

\bibitem[Dhomkar \em{et~al.}(2016)Dhomkar, Henshaw, Jayakumar, and
 Meriles]{Dhomkar2016}
Dhomkar, S.; Henshaw, J.; Jayakumar, H.; Meriles, C.
\newblock {Long-term data storage in diamond}.
\newblock {\em Sci. Adv.} {\bf 2016}, {\em 2},~e1600911.

\bibitem[Aslam \em{et~al.}(2013)Aslam, Waldherr, Neumann, Jelezko, and
 Wrachtrup]{Aslam2013}
Aslam, N.; Waldherr, G.; Neumann, P.; Jelezko, F.; Wrachtrup, J.
\newblock {Photo-induced ionization dynamics of the nitrogen vacancy defect in
 diamond investigated by single-shot charge state detection}.
\newblock {\em New J. Phys.} {\bf 2013}, {\em 15},~013064.

\bibitem[Jaskula \em{et~al.}(2017)Jaskula, Shields, Bauch, Lukin, and
 Trifonov]{Jaskula}
Jaskula, J.; Shields, B.J.; Bauch, E.; Lukin, M.D.; Trifonov, A.S.
\newblock {Improved quantum sensing with a single solid- state spin via
 spin-to-charge conversion} \emph{arXiv} {\bf 2017}, arxiv:1711.02023.

\bibitem[Ariyaratne \em{et~al.}(2018)Ariyaratne, Bluvstein, Myers, and
 Jayich]{Ariyaratne2018}
Ariyaratne, A.; Bluvstein, D.; Myers, B.A.; Jayich, A.C.B.
\newblock {Nanoscale electrical conductivity imaging using a nitrogen-vacancy
 center in diamond}.
\newblock {\em Nat. Commun.} {\bf 2018}, {\em 9},~2406.

\bibitem[Bourgeois \em{et~al.}(2015)Bourgeois, Jarmola, Siyushev, Gulka, Hruby,
 Jelezko, Budker, and Nesladek]{Bourgeois2015}
Bourgeois, E.; Jarmola, A.; Siyushev, P.; Gulka, M.; Hruby, J.; Jelezko, F.;
 Budker, D.; Nesladek, M.
\newblock {Photoelectric detection of electron spin resonance of
 nitrogen-vacancy centres in diamond}.
\newblock {\em \mbox{Nat. Commun.}} {\bf 2015}, {\em 6},~8577.

\bibitem[Brenneis \em{et~al.}(2015)Brenneis, Gaudreau, Seifert, Karl, Brandt,
 Huebl, Garrido, L., and Holleitner]{Brenneis2015}
Brenneis, A.; Gaudreau, L.; Seifert, M.; Karl, H.; Brandt, M.S.; Huebl, H.;
 Garrido, J.A.; Koppens, F.H.; Holleitner, A.W.
\newblock {Ultrafast electronic readout of diamond nitrogen-vacancy centres
 coupled to graphene}.
\newblock {\em \mbox{Nat. Nanotechnol.}} {\bf 2015}, {\em 10},~135--139.

\bibitem[Bourgeois \em{et~al.}(2017)Bourgeois, Londero, Buczak, Hruby, Gulka,
 Balasubramaniam, Wachter, Stursa, Dobes, Aumayr, Trupke, Gali, and
 Nesladek]{Bourgeois2017}
Bourgeois, E.; Londero, E.; Buczak, K.; Hruby, J.; Gulka, M.; Balasubramaniam,
 Y.; Wachter, G.; Stursa, J.; Dobes, K.; Aumayr, F.; et al.
\newblock {Enhanced photoelectric detection of NV magnetic resonances in
 diamond under dual-beam excitation}.
\newblock {\em Phys. Rev. B} {\bf 2017}, {\em 95},~41402.

\bibitem[Hrubesch \em{et~al.}(2017)Hrubesch, Braunbeck, Stutzmann, Reinhard,
 and Brandt]{Hrubesch2017}
Hrubesch, F.M.; Braunbeck, G.; Stutzmann, M.; Reinhard, F.; Brandt, M.S.
\newblock {Efficient Electrical Spin Readout of NV$^-$ Centers in Diamond}.
\newblock {\em Phys. Rev. Lett.} {\bf 2017}, {\em 118},~037601.

\bibitem[Gulka \em{et~al.}(2017)Gulka, Bourgeois, Hruby, Siyushev, Wachter,
 Aumayr, Hemmer, Gali, Jelezko, Trupke, and Nesladek]{Gulka2017}
Gulka, M.; Bourgeois, E.; Hruby, J.; Siyushev, P.; Wachter, G.; Aumayr, F.;
 Hemmer, P.R.; Gali, A.; Jelezko, F.; Trupke, M.; Nesladek, M.
\newblock {Pulsed Photoelectric Coherent Manipulation and Detection of NV$^-$
 Center Spins in Diamond}.
\newblock {\em Phys. Rev. Appl.} {\bf 2017}, {\em 7},~044032.

\bibitem[Heremans \em{et~al.}(2009)Heremans, Fuchs, Wang, Hanson, and
 Awschalom]{Heremans2009}
Heremans, F.J.; Fuchs, G.D.; Wang, C.F.; Hanson, R.; Awschalom, D.D.
\newblock {Generation and transport of photoexcited electrons in single-crystal
 diamond}.
\newblock {\em Appl. Phys. Lett.} {\bf 2009}, {\em 94},~7--10.

\bibitem[Bourdeauducq \em{et~al.}(2018)Bourdeauducq, Whitequark, J{\"{o}}rdens,
 Sionneau, Enjoy-digital, Cjbe, JBoulder, Hartytp, Slichter, Mntng, Nadlinger,
 R-srinivas, Britton, Smith, Kemstevens, Held, and Leibrandt]{artiq}
Bourdeauducq, S.; Whitequark.; J{\"{o}}rdens, R.; Sionneau, Y.; Enjoy-digital.;
 Cjbe.; JBoulder.; Hartytp.; Slichter, D.; Mntng.; Nadlinger, D.; R-srinivas.;
 Britton, J.; Smith, Z.; Kemstevens.; Held, F.; Leibrandt, D.
\newblock m-labs/artiq: 3.6 {\bf 2018}.

\bibitem[Binder \em{et~al.}(2017)Binder, Stark, Tomek, Scheuer, Frank, Jahnke,
 Müller, Schmitt, Metsch, Unden, Gehring, Huck, Andersen, Rogers, and
 Jelezko]{Qudi2017}
Binder, J.M.; Stark, A.; Tomek, N.; Scheuer, J.; Frank, F.; Jahnke, K.D.;
 Müller, C.; Schmitt, S.; Metsch, M.H.; Unden, T.; et al.
\newblock Qudi: A modular python suite for experiment control and data
 processing.
\newblock {\em SoftwareX} {\bf 2017}, {\em 6},~85--90.

\bibitem[Lim \em{et~al.}(2007)Lim, Mar, Matheu, Derkacs, and Yu]{Lim2007}
Lim, S.H.; Mar, W.; Matheu, P.; Derkacs, D.; Yu, E.T.
\newblock {Photocurrent spectroscopy of optical absorption enhancement in
 silicon photodiodes via scattering from surface plasmon polaritons in gold
 nanoparticles}.
\newblock {\em \mbox{J. Appl. Phys.}} {\bf 2007}, {\em 101},~104309.

\bibitem[{Tse-Luen Wee} \em{et~al.}(2007){Tse-Luen Wee}, Tzeng, Han, Chang,
 Fann, Hsu, Chen, and Yu]{Tse-LuenWee2007}
{Tse-Luen Wee}.; Tzeng, Y.K.; Han, C.C.; Chang, H.C.; Fann, W.; Hsu, J.H.;
 Chen, K.M.; Yu, Y.C.
\newblock {Two-photon Excited Fluorescence of Nitrogen-Vacancy Centers in
 Proton-Irradiated Type lb Diamond}.
\newblock {\em J. Phys. Chem. A} {\bf 2007}, {\em
 111},~9379--9386.

\bibitem[Bogdanov \em{et~al.}(2018)Bogdanov, Shalaginov, Lagutchev, Chiang,
 Shah, Baburin, Ryzhikov, Rodionov, Kildishev, Boltasseva, and
 Shalaev]{Bogdanov2017a}
Bogdanov, S.I.; Shalaginov, M.Y.; Lagutchev, A.S.; Chiang, C.C.; Shah, D.;
 Baburin, A.S.; Ryzhikov, I.A.; Rodionov, I.A.; Kildishev, A.V.; Boltasseva,
 A.; et al.
\newblock Ultrabright Room-Temperature Sub-Nanosecond Emission from Single
 Nitrogen-Vacancy Centers Coupled to Nanopatch Antennas.
\newblock {\em Nano Lett.} {\bf 2018}, {\em 18}, 4837--4844.

\end{thebibliography}

\end{document}